\documentclass[10pt,a4paper]{article}

\usepackage{jheppub}

\usepackage{color}
\usepackage{colordvi}
\usepackage{changes}
\definechangesauthor{SM}{red}

\usepackage{epsfig,epsf}
\usepackage{amsmath}
\usepackage{amsthm}
\usepackage{amsfonts}
\usepackage{amssymb}
\usepackage{dsfont}
\usepackage{epstopdf}
\usepackage{multirow}

\usepackage{rotating}

\usepackage{marvosym}

\usepackage{subcaption}
\usepackage{array}   
\newcolumntype{C}{>{$}c<{$}}

\usepackage{slashed}

\usepackage[active]{srcltx}

\def\II{\hbox{{1}\kern-.25em\hbox{l}}}

\DeclareMathOperator{\Li}{Li}

\def\II{\hbox{{1}\kern-.25em\hbox{l}}}

\title{
\begin{flushright}
{\large \textnormal{DESY 17-045}}\\[2mm]
\end{flushright}
Three-loop evolution equation for flavor-nonsinglet operators in off-forward kinematics}

\author[a]{V. M. Braun,}
\author[b,a]{A. N. Manashov,}
\author[b]{S. Moch}
\author[a]{and M. Strohmaier}

\affiliation[a]{
   Institut f\"ur Theoretische Physik, Universit\"at
   Regensburg \\ D-93040 Regensburg, Germany}
\affiliation[b]{
   Institut f\"ur Theoretische Physik, Universit\"at Hamburg\\
   D-22761 Hamburg, Germany}

\emailAdd{alexander.manashov@desy.de}

\emailAdd{sven-olaf.moch@desy.de}

\emailAdd{matthias.strohmaier@ur.de}

\abstract{
Using the approach based on conformal symmetry we calculate the three-loop (NNLO) contribution to the
evolution equation for flavor-nonsinglet leading twist operators in the
$\overline{\text{MS}}$ scheme. 
The explicit expression for the three-loop kernel is derived for the corresponding
light-ray operator in coordinate space. The expansion in local operators
is performed and explicit results 
are given for the matrix of the anomalous dimensions for the operators up to
seven covariant derivatives.
The results are directly applicable  to the renormalization of the pion light-cone distribution amplitude
and flavor-nonsinglet generalized parton distributions.
       }


\begin{document}
\maketitle

\section{Introduction}

The remarkable progress in experimental techniques in the past two decades has
provided a fresh impetus to
the study of hard exclusive and semi-inclusive reactions with identified particles
in the final state.
Such processes are interesting as they allow one to access 
the hadron structure on a much more detailed
level as compared to totally inclusive reactions. A (probably still distant) major goal is to understand
the full three-dimensional proton structure by ``holographic imaging'' of quark and gluon distributions
in transverse distance and momentum spaces.
The related experiments have become a prominent part of the research program at all major existing
and planned accelerator facilities, e.g. the Electron Ion Collider (EIC)~\cite{Boer:2011fh}.

The relevant nonperturbative input in such processes in many cases involves operator matrix
elements between states with different momenta, dubbed generalized parton distributions (GPDs),
or vacuum-to-hadron matrix elements related to light-front hadron wave functions at small trans-
verse separations, the distribution amplitudes (DAs).
The scale-dependence of such distributions is governed by the renormalization group (RG) equations
for the corresponding operators, where, in 
contrast to standard parton densities,
mixing with the operators involving total derivatives has to be taken into account.
Going over to local operators one has to deal with a triangular mixing matrix where the
diagonal entries are the anomalous dimensions, the same as in deep-inelastic scattering,
but the off-diagonal elements require a separate calculation.

The projected very high accuracy of future experimental data,
e.g. on the Deeply Virtual Compton Scattering (DVCS) at the JLAB 12 GeV upgrade~\cite{Dudek:2012vr} and the EIC,
and the $\gamma^*\to\pi\gamma$ transition form factor at Belle II at KEK~\cite{Abe:2010gxa}, has to be matched by the
increasing theoretical precision; in 
the ideal case one would like to reach the same level of
accuracy as in inclusive reactions. The NNLO (three-loop) analysis of parton distributions and
fragmentation functions is becoming the standard in this field~\cite{Accardi:2016ndt}, so that the NNLO evolution
equations for off-forward distributions are appearing on the agenda.

In this work we derive the explicit expression for the three-loop contribution to the flavor-nonsinglet
evolution kernel in the so-called light-ray operator representation. This kernel can be converted to the
evolution equation for the GPDs by a Fourier transformation, whereas its expansion at small distances
provides one with the matrix of the anomalous dimensions for local leading twist operators.
In the latter form, our results are directly relevant for the lattice calculations of
pion DAs in which case the uncertainty due to the conversion of lattice results to the
$\overline{\text{MS}}$ scheme 
currently proves to be one of the dominant sources of the error~\cite{Braun:2015axa}.
The three-loop (NNLO) anomalous dimensions of the leading-twist operators
are known for about a decade~\cite{Moch:2004pa}, however, a direct calculation of the
missing off-diagonal terms in the mixing matrix to the same precision is quite challenging.

Conformal symmetry of the QCD Lagrangian allows one to restore 
the nondiagonal entries in the mixing matrix
and, hence, full evolution kernels at a given order of perturbation theory from the calculation of
the special conformal anomaly at one order less~\cite{Mueller:1991gd}.
This result was used to calculate the complete two-loop mixing matrix for
twist-two operators in QCD~\cite{Mueller:1993hg,Mueller:1997ak,Belitsky:1997rh}, and 
to derive the two-loop evolution kernels for the GPDs~\cite{Belitsky:1998vj,Belitsky:1999hf,Belitsky:1998gc}.

In Ref.~\cite{Braun:2013tva} we have proposed 
to use a somewhat different technique to implement the same
idea. Instead of studying conformal symmetry {\it breaking} in the physical
theory~\cite{Mueller:1993hg,Mueller:1997ak,Belitsky:1997rh} we suggest to make use of the
{\it exact} conformal symmetry of large-$n_f$ QCD in $d=4-2\epsilon$ dimensions at critical coupling.
Due to specifics of the minimal subtraction scheme ($\overline{\text{MS}}$)
the renormalization group equations in 
the physical four-dimensional theory inherit a conformal symmetry
so that the evolution kernel commutes with the generators of conformal transformations.
This symmetry is exact, however, the generators are modified by quantum corrections and differ from their canonical form.
The consistency relations that follow from the conformal algebra can be used in order to restore  the
$\ell$-loop off-forward kernel from the $\ell$-loop anomalous dimensions and the $(\ell-1)$-loop
result for the deformation of the generators, which is equivalent to the statement in
Ref.~\cite{Mueller:1991gd}.

Exact conformal symmetry of modified QCD allows one to use algebraic group-theory methods to
resolve the constraints on the operator mixing and also suggests the optimal representation for the results
in terms of light-ray operators. In this way one avoids the need to restore the evolution kernels
from the results for local operators, which is not straightforward.
This modified approach was tested in~\cite{Braun:2013tva} on several examples to two- and
three-loop accuracy for scalar theories, and in~\cite{Braun:2014vba} for flavor-nonsinglet operators in QCD
to two-loop accuracy. As a major step towards the NNLO calculation, in~\cite{Braun:2016qlg} we have calculated
the two-loop quantum correction to the generator of special conformal transformations.
In this work we use this result to obtain the three-loop (NNLO) evolution equation for flavor-nonsinglet
leading twist operators in the light-ray operator representation in the $\overline{\text{MS}}$ scheme.
The relation to the representation in terms of local operators~\cite{Mueller:1991gd,Mueller:1993hg,Mueller:1997ak}
is worked out in detail and explicit results are given for the matrix of the anomalous dimensions for the
operators with up to seven covariant derivatives.
Our results are directly applicable e.g. to the studies of the pion light-cone DA and flavor-nonsinglet GPDs.

The presentation is organized as follows. Sect.~\ref{sec:general} is introductory, it contains
a very short general description of the light-ray operator formalism and
the conformal algebra.
In this section we also explain our notation and conventions. In Sect.~\ref{sec:similarity} we show that
the contributions to the evolution kernel due to 
the conformal anomaly can be isolated by a similarity
transformation. As the result, the evolution kernel can be written as a sum of several contributions with
a simpler structure. This construction is similar in spirit to the ``conformal scheme'' of
Refs.~\cite{Mueller:1993hg,Mueller:1997ak,Belitsky:1997rh}. We find that the remaining (canonically)
$SL(2)$-invariant part of the three-loop kernel satisfies the reciprocity relation
\cite{Dokshitzer:2005bf,Basso:2006nk,Alday:2015eya,Alday:2015ewa}, discussed in Sect.~\ref{sec:reciprocity}.
The explicit construction of the invariant kernel is presented in Sect.~\ref{sec:H3inv}.
We provide analytic expressions for the terms that correspond to
the leading asymptotic behavior at small and large Bjorken $x$, and a simple parametrization for the remainder
that has sufficient accuracy for all potential applications. In Sect.~\ref{sec:localOPE} we explain how our results
for the renormalization of light-ray operators can be translated into anomalous dimension matrices for local
operators. In this way also the formal relation to the results in~\cite{Mueller:1991gd,Mueller:1993hg,Mueller:1997ak}
is established. The final Sect.~\ref{sec:summary} is reserved for a summary and outlook.
The paper also contains several Appendices where we collect the analytic expressions for the kernels.

\section{Evolution equations for light-ray operators }
\label{sec:general}

A renormalized light-ray operator,
\begin{align}
 [\mathcal{O}](x;z_1,z_2) = Z \mathcal{O}(x;z_1,z_2) = Z \bar q(x+z_1n)\slashed{n} q(x+z_2n),
\end{align}
where the Wilson line is implied between the quark fields on the light-cone,
is defined as the generating function for renormalized local operators:
\begin{eqnarray}
 [\mathcal{O}](x;z_1,z_2) &\equiv& \sum_{m,k} \frac{z_1^m z_2^k}{m!k!}
\left[\bar q(x) (\stackrel{\leftarrow}{D}\!\cdot n)^m \slashed{n} (n\cdot\! \stackrel{\rightarrow}{D})^k q(x)\right],
\label{LRO}
\end{eqnarray}
where $D_\mu =\partial_\mu -i g A_\mu$ is the covariant derivative.
Here and below we
use square brackets to denote renormalized composite
operators (in a minimal subtraction scheme).
Due to Poincare invariance in most situations one can put $x=0$ without loss of generality;
we will therefore often drop the $x$ dependence
and write $$\mathcal{O}(z_1,z_2) \equiv \mathcal{O}(0; z_1,z_2).$$
The renormalization factor $Z$ is an integral operator in $z_1,z_2$ which is given by a series in~$1/\epsilon$, $d=4-2\epsilon$,
\begin{align}\label{Zfactor}
Z=1+\sum_{k=0}^\infty  \frac1{\epsilon^k} Z_k(a)\,, && Z_k(a)=\sum_{\ell=k}^\infty a^\ell Z_{k}^{(\ell)}\,.
\end{align}
The RG equation for the light-ray operator $[\mathcal{O}]$ takes the form
\begin{align}\label{RGO}
\Big(M{\partial_M}+\beta(a)\partial_a +\mathbb{H}(a)\Big)[\mathcal{O}](x;z_1,z_2)=0\,,
\end{align}
where $M$ is the renormalization scale,
\begin{align}
a &= \frac{\alpha_s}{4\pi}\,,
\qquad
\beta(a)= M\frac{da}{d M} =-2a\big(\epsilon + a\beta_0+a^2\beta_1+\ldots\big)=-2a(\epsilon+ \bar \beta(a))\,
\label{beta}
\end{align}
with
\begin{align}
\beta_0 & =\frac{11}3N_c-\frac23n_f\,, && \beta_1 = \frac23\left[17C_A^2-5C_A n_f-3C_F n_f\right]\,.
\end{align}
$\mathbb{H}(a)$ is an integral operator acting on the light-cone coordinates of the fields,
which has a perturbative expansion
\begin{align}
   \mathbb{H}(a) &= a\,\mathbb{H}^{(1)} + a^2\,\mathbb{H}^{(2)} + a^3\,\mathbb{H}^{(3)} + \ldots
\end{align}
 It is related to the renormalization factor~\eqref{Zfactor} as follows
\begin{align}\label{H-Z}
 \mathbb{H}(a)=-M\frac{d}{d M}\mathbb{Z} \mathbb{Z}^{-1}=2\gamma_q(a)+2\sum_{\ell=1}^\infty  \ell\,a^\ell Z_1^{(\ell)}\,,
\end{align}
where $\mathbb{Z}=ZZ_q^{-2}$; $Z_q$ is the quark wave function renormalization factor and
$\gamma_q= M\partial_M \ln Z_q$ the quark anomalous dimension. 
The QCD $\beta$-function $\beta(a)$ and $\gamma_q$ are known to
$\mathcal{O}(a^5)$~\cite{Baikov:2014qja,Baikov:2016tgj,Luthe:2016xec,Luthe:2017ttc,Baikov:2017ujl}.

The evolution operator can be written as~\cite{Balitsky:1987bk}
\begin{align}
 \mathbb{H}(a)[\mathcal{O}](z_1,z_2) = \int_0^1 \!d\alpha\int_0^1\! d\beta\, h(\alpha,\beta)\, [\mathcal{O}](z_{12}^\alpha,z_{21}^\beta)\,,
\label{hkernel}
\end{align}
where
\begin{align}
z_{12}^\alpha =  z_1\bar\alpha+z_2\alpha && \bar\alpha=1-\alpha\,,
\label{z12alpha}
\end{align}
and $h(\alpha,\beta)=a\, h^{(1)}(\alpha,\beta)+a^2 h^{(2)}(\alpha,\beta)+\ldots$ is a certain weight function (evolution kernel).

It is easy to see~\cite{Braun:2013tva} that translation-invariant polynomials $(z_1-z_2)^N$ are eigenfunctions
of the evolution kernel,
\begin{align}
    \mathbb{H}(a) z_{12}^N &=  \gamma_N(a) \,  z_{12}^N \qquad\qquad z_{12} = z_1-z_2\,.
\label{adim1}
\end{align}
The eigenvalues $\gamma_N(a)$ correspond to moments of the evolution kernel in the representation~\eqref{hkernel},
\begin{align}
 \gamma_N &= \int_0^1 \!d\alpha\int_0^1\! d\beta\, (1-\alpha-\beta)^N h(\alpha,\beta) = a \gamma_N^{(1)} + a^2 \gamma_N^{(2)} + a^3 \gamma_N^{(3)} +\ldots\,.
\label{hmoments}
\end{align}
They define the anomalous dimensions of leading-twist local operators in Eq.~\eqref{LRO} where $N=m+k$ is the total number of covariant
derivatives acting either on the quark or the antiquark field.
The corresponding mixing matrix in the Gegenbauer polynomial basis is constructed in Sect.~\ref{sec:localOPE}.

The leading-order (LO) result for the evolution kernel in this representation reads~\cite{Balitsky:1987bk}:
\begin{align}
\mathbb{H}^{(1)}f(z_1,z_2)&=4C_F\biggl\{
\int_0^1d\alpha\frac{\bar\alpha}{\alpha}\Big[2f(z_1,z_2)-f(z_{12}^\alpha,z_2)-f(z_1,z_{21}^\alpha)\Big]
\notag\\
&\quad
-\int_0^1d\alpha\int_0^{\bar\alpha}d\beta \, f(z_{12}^\alpha,z_{21}^\beta)
+\frac12 f(z_1,z_2)
\biggr\}\,.
\label{Honeloop}
\end{align}
The expression in Eq.~(\ref{Honeloop}) gives rise to all classical
leading-order (LO) QCD evolution equations: the DGLAP equation for parton distributions,
the ERBL equation for the meson light-cone DAs, and the general evolution equation for GPDs.

The LO evolution kernel $\mathbb{H}^{(1)}$ commutes with the (canonical) generators
of collinear conformal transformations
\begin{align}\label{canon}
S^{(0)}_-&=-\partial_{z_1}-\partial_{z_2}\,,
\notag\\
S^{(0)}_0&=z_1\partial_{z_1}+z_2\partial_{z_2}+2,
\notag\\
S^{(0)}_+&=z_1^2\partial_{z_1}+z_2^2\partial_{z_2}+2(z_1+z_2)\,
\end{align}
which satisfy the usual $SL(2)$ algebra
\begin{align}\label{sl2-comm}
{}[S_0,S_{\pm}]=\pm S_{\pm}\,, && {}[S_{+},S_-]= 2S_0\,.
\end{align}
It can be shown that as a consequence of the commutation relations $[\mathbb{H}^{(1)},S^{(0)}_\alpha]=0$
the corresponding kernel $h^{(1)}(\alpha,\beta)$ is effectively a function of one variable $\tau$ called the conformal
ratio~\cite{Braun:1999te}
\begin{align}\label{hinv}
h^{(1)} (\alpha,\beta) = \bar h (\tau)\,, && \tau = \frac{\alpha\beta}{\bar\alpha\bar\beta}\,,
\end{align}
up to trivial terms $\sim\delta(\alpha)\delta(\beta)$ that correspond to the unit operator.
This function can easily be reconstructed from its moments (\ref{hmoments}), alias from the anomalous dimensions.

Indeed, it is easy to verify that the result in Eq.~\eqref{Honeloop} can be
written in the following, remarkably simple form~\cite{Braun:1999te}
\begin{align}
      h^{(1)}(\alpha,\beta) = -4 C_F\left[\delta_+(\tau) + \theta(1-\tau)-\frac12\delta(\alpha)\delta(\beta)\right],
\label{QCD-LO}
\end{align}
where the regularized $\delta$-function, $\delta_+(\tau)$, is defined as
\begin{align}
\int d\alpha d\beta\, \delta_+(\tau)f(z_{12}^\alpha,z_{21}^\beta)&\equiv\int_0^1 d\alpha\int_0^{1} d\beta\, \delta(\tau)
\Big[f(z_{12}^\alpha,z_{21}^\beta)-f(z_1,z_2)\Big]
\notag\\
&=-
\int_0^1 d\alpha \frac{\bar \alpha}{\alpha}\Big[2f(z_1,z_2)-f(z_{12}^\alpha,z_2)-f(z_1,z_{21}^{\alpha})\Big].
\label{delta-plus}
\end{align}

Beyond the LO this property is lost. However, the evolution kernels for leading twist operators
in minimal subtraction schemes retain \emph{exact} conformal symmetry. Indeed, the renormalization
factors for composite operators in this scheme do not depend on $\epsilon$ by construction.
As a consequence, the anomalous dimension matrices in QCD in four dimensions are exactly the
same as in QCD in $d=4-2\epsilon$ dimensions that enjoys conformal symmetry for the specially chosen
``critical'' value of the coupling~\cite{Braun:2013tva,Braun:2014vba,Braun:2016qlg}.
The precise statement is that the QCD evolution kernel $\mathbb{H}(a)$ commutes with three operators
\begin{align}
  [\mathbb{H}(a),S_+]~=~[\mathbb{H}(a),S_-] ~=~ [\mathbb{H}(a),S_0] ~=~ 0
\end{align}
that satisfy the $SL(2)$ algebra~\eqref{sl2-comm}.
These operators can be constructed as the generators of collinear conformal transformations in
the $4-2\epsilon$-dimensional QCD at the critical point and have the following structure~\cite{Braun:2013tva,Braun:2014vba,Braun:2016qlg}:
\begin{subequations}
\label{exactS}
\begin{align}\label{translation}
   S_- &= S_-^{(0)}\,,
\\
\label{dilatation}
   S_0\, &= S_0^{(0)} +\Delta S_0 ~=~ S_0^{(0)}  +\left(\bar\beta(a)+ \frac12\mathbb{H}(a)\right)\,,
\\
   S_+ &=  S_+^{(0)} + \Delta S_+ ~=~  S_+^{(0)} + (z_1+z_2)\left(\bar\beta(a)+ \frac12  \mathbb{H}(a)\right)
+   (z_1-z_2)\,\Delta(a)\,,
\label{special-conformal}
\end{align}
\end{subequations}
where $\bar\beta(a)$ is the QCD $\beta$-function \eqref{beta} and $S_\alpha^{(0)}$ are the canonical generators~\eqref{canon}.

Note that the generator $S_-$ corresponds to translations along the light cone and does not receive any corrections
as compared to its canonical expression, $S_-^{(0)}$. The generator
$S_0$ corresponds to dilatations; its modification in interacting theory $\Delta S_0 = S_0-S_0^{(0)}$ can be related to the
evolution kernel $\mathbb{H}(a)$ from general considerations~\cite{Braun:2013tva}. Finally $S_+$ is the generator of
special conformal transformations and Eq.~\eqref{special-conformal} is the most general expression consistent with
the commutation relations~\eqref{sl2-comm}. To see that, note that $\Delta S_+ = S_+ -S_+^{(0)}$ must have the same
canonical dimension $[\text{mass}]^{-1}$ as $S_+^{(0)}$, meaning that $[S_0^{(0)},\Delta S_+]=\Delta S_+$. Thus we
can write $\Delta S_+ = (z_1+z_2) \Delta_1 + (z_1-z_2) \Delta_2$ where $[S_0^{(0)}, \Delta_{1,2}]=0$. Plugging this
expression in the commutation relation $[S_+,S_-]=2S_0$ one obtains
$\Delta_1=\Delta S_0$ and $[S_-,\Delta_2]=0$. Changing notation $\Delta_2 \mapsto \Delta$ we arrive at the
expression given in  Eq.~\eqref{special-conformal}.

Using \eqref{exactS} in the commutation relation $[S_0,S_+]=S_+$, or equivalently $[\mathbb{H}(a), S_+]=0$, results in
\begin{align}\label{HDelta}
\big[S_+^{(0)},\mathbb{H}(a)\big]= -\big[\Delta S_+, \mathbb{H}(a)\big]=
\big[\mathbb{H}(a),z_1+z_2\big]\left(\bar\beta(a)+ \frac12\mathbb{H}(a)\right)+\big[\mathbb{H}(a),z_{12} \Delta(a)\big].
\end{align}
If $\mathbb{H}(a)$ is known, this equation can be used to find $\Delta(a)$ and in this way
construct the $SL(2)$ generators that commute with the evolution kernel in a theory with broken conformal symmetry.
The main point is, however, that $\Delta(a)$ can be calculated independently from the analysis of the
conformal Ward identity~\cite{Mueller:1991gd,Belitsky:1998gc,Braun:2016qlg}.
In this way Eq.~\eqref{HDelta} can be used to calculate the non-invariant part of the evolution kernel
with respect to the canonical generators $S_{\pm,0}^{(0)}$.

Indeed, expanding the kernels in a power series in coupling constant
\begin{align}
\mathbb{H}(a) =a \mathbb{H}^{(1)}+ a^2 \mathbb{H}^{(2)}+a^3 \mathbb{H}^{(3)}+\ldots,
&&
\Delta(a) =a \Delta^{(1)}+ a^2 \Delta^{(2)} +\ldots
\end{align}
one obtains from \eqref{HDelta} a nested set of equations~\cite{Braun:2013tva}
\begin{subequations}\label{HHD}
\begin{align}
[S_+^{(0)},\mathbb{H}^{(1)}] & = 0\,,
\label{S0H1a}\\
[S_+^{(0)},\mathbb{H}^{(2)}] & =
\big[\mathbb{H}^{(1)},z_1+z_2\big]\left(\beta_0+\frac12 \mathbb{H}^{(1)}\right)+\big[\mathbb{H}^{(1)},z_{12} \Delta^{(1)}\big]\,,
\label{S0H2a}\\
[S_+^{(0)},\mathbb{H}^{(3)}] & =
\big[\mathbb{H}^{(1)},z_1+z_2\big]\left(\beta_1+ \frac12\mathbb{H}^{(2)}\right)+\big[\mathbb{H}^{(2)},z_1
+z_2\big]\left(\beta_0+ \frac12\mathbb{H}^{(1)}\right)
\notag\\
&\quad
+ \big[\mathbb{H}^{(2)},z_{12} \Delta^{(1)}\big]+ \big[\mathbb{H}^{(1)},z_{12} \Delta^{(2)}\big]\,,
\label{S0H3a}
\end{align}
\end{subequations}
so that the commutator $[S_+^{(0)},\mathbb{H}^{(\ell)}]$ is expressed in terms of the lower order kernels, $\mathbb{H}^{(k)}$ and
$\Delta^{(k)}$ with $k \le \ell-1$.

The first of them, Eq.~\eqref{S0H1a}, is the usual statement that the LO evolution kernel commutes with canonical generators of the conformal
transformation. As a consequence, the corresponding kernel $h^{(1)}(\alpha,\beta)$ can be written as a function of the conformal ratio
\eqref{hinv} and restored from the spectrum of LO anomalous dimensions. The result is presented in Eqs.~\eqref{Honeloop}, \eqref{QCD-LO}.

The second equation, Eq.~\eqref{S0H2a}, is, technically,  a first-order inhomogeneous differential equation on the
NLO evolution kernel $\mathbb{H}^{(2)}$. To solve this equation one needs to find a particular solution with the given
inhomogeneity (the expression on the r.h.s.), and add a solution of the homogeneous equation $[S_+^{(0)},\mathbb{H}^{(2)}]$ such 
that the sum reproduces
the known NLO anomalous dimensions. This calculation was done in Ref.~\cite{Braun:2014vba} and the final expression for $\mathbb{H}^{(2)}$
is reproduced in a somewhat different form below in App.~\ref{app:H2loop}.

In this work we solve Eq.~\eqref{S0H3a} and in this way calculate the three-loop (NNLO) evolution kernel $\mathbb{H}^{(3)}$.
The main input in this calculation is provided by the recent result for the two-loop conformal anomaly $\Delta^{(2)}$~\cite{Braun:2016qlg}.
Since the algebraic structure of the expressions at the three-loop level is quite complicated, we separate the calculation
in several steps in order to disentangle contributions of different origin.
The basic idea is to simplify the structure as much as possible by separating parts of the three-loop kernel that can be written
as a product of simpler kernels.

\section{Similarity transformation}\label{sec:similarity}

The symmetry generators $S_\alpha$ in a generic interacting theory \eqref{exactS} involve the evolution kernel $\mathbb{H}(a)$
and additional contributions $\Delta(a)$ due to the conformal anomaly.
These two terms can be separated by a similarity transformation
\begin{align}
    \mathbb{H} = \mathrm{U}^{-1}\, \mathbf{H}\, \mathrm{U}\,, \qquad S_{\pm,0} =  \mathrm{U}^{-1}\, \mathbf{S}_{\pm,0}\, \mathrm{U}\,.
\label{similarity1}
\end{align}
Note that $\mathbb{H}$ and $\mathbf{H}$ obviously have the same eigenvalues (anomalous dimensions).
Going over to the ``boldface'' operators can be thought of as a change of the
renormalization scheme,
\begin{align}
{}[\mathcal{O}(z_1,z_2)]_U= \mathrm{U}\, [\mathcal{O}(z_1,z_2)]_{\overline{\text{MS}}}.
\end{align}
The ``rotated'' light-ray operator $[\mathcal{O}(z_1,z_2)]_U$ satisfies the RG equation
\begin{align}\label{RGOU}
\Big(M{\partial_M}+ \beta(a)\partial_a +  \mathbf{H}(a) 
- {\beta}(a)\partial_a \mathrm{U}\cdot \mathrm{U}^{-1} 
\Big)[\mathcal{O}(z_1,z_2)]_U=0\,.
\end{align}
 Looking for the operator $\mathrm{U}$ in the form
\begin{align}
   \mathrm{U} = e^{\mathbb{X}}\,, \qquad \mathbb{X}(a)  =  a \mathbb{X}^{(1)}+ a^2 \mathbb{X}^{(2)}+\ldots\,,
\label{similarity2}
\end{align}
we require that the ``boldface'' generators do not include conformal anomaly terms,
\begin{subequations}
\label{Sbold}
\begin{align}\label{translationbold}
   \mathbf{S}_- &= {S}_-^{(0)}\,,
\\
\label{dilatationbold}
   \mathbf{S}_0\, &= {S}_0^{(0)} +\Delta \mathbf{S}_0 ~=~ {S}_0^{(0)}  +\left(\bar\beta(a)+ \frac12\mathbf{H}(a)\right)\,,
\\
   \mathbf{S}_+ &=  {S}_+^{(0)} + \Delta \mathbf{S}_+ ~=~ {S}_+^{(0)} + (z_1+z_2)\left(\bar\beta(a)+ \frac12  \mathbf{H}(a)\right)\,.
\label{special-conformalbold}
\end{align}
\end{subequations}
With this choice the generators $\mathbf{S}_\alpha$ on the subspace of the eigenfunctions of the operator $\mathbf{H}$ 
with a given anomalous dimension $\gamma_N$  take the canonical form with shifted conformal spin $j=1 \to 1+\frac12 \bar\beta(a)+\frac14\gamma_N(a)$
so that the eigenfunctions of  $\mathbf{H}$ can be constructed explicitly. The evolution equation in this form~\eqref{RGOU}
still contains, however, an extra term 
$\beta(a)\partial_a \mathrm{U}\cdot \mathrm{U}^{-1}$
and is not diagonalized. 
   
 Since the evolution kernel commutes with the canonical generators $S^{(0)}_-$ and $S^{(0)}_0$ we can assume  that
$\mathbb{X}^{(k)}$ commute with the same generators as well,
\begin{align}\label{X12}
[S_-^{(0)},\mathbb{X}^{(k)}]=[S_0^{(0)},\mathbb{X}^{(k)}]=0\,,
\end{align}
whereas comparing Eqs.~\eqref{special-conformal} and \eqref{special-conformalbold} yields the following set of equations:
\begin{subequations}\label{X12def}
\begin{align}
\big[S_+^{(0)}, \mathbb{X}^{(1)}\big] &= z_{12} \Delta^{(1)}\,,
\label{X1def}\\
\big[S_+^{(0)}, \mathbb{X}^{(2)}\big] &= z_{12} \Delta^{(2)}+\Big[\mathbb{X}^{(1)},z_1+z_2\Big]\left(\beta_0+\frac12\mathbb{H}^{(1)}\right)
+\frac12\Big[\mathbb{X}^{(1)},z_{12}\Delta^{(1)}\Big].
\label{X2def}
\end{align}
\end{subequations}
These equations can be used to fix $\mathbb{X}^{(1)}$ and $\mathbb{X}^{(2)}$ from the known one- and two-loop
expressions for the conformal anomaly, $\Delta^{(1)}$ and $\Delta^{(2)}$~\cite{Braun:2014vba,Braun:2016qlg}, up to
$SL(2)$ (canonically) invariant terms. In other words, the transformation $\mathrm{U}$ that brings the conformal generators
to the form \eqref{Sbold} is not unique; there is some freedom and we specify our choice later on.
The one-loop result, $\mathbb{X}^{(1)}$, turns out to be rather simple whereas the two-loop operator, $\mathbb{X}^{(2)}$, is
considerably more involved. Explicit expressions are presented in App.~\ref{app:X}.

The rotated, ``boldface'' evolution kernels satisfy a simpler set of equations as compared to Eqs.~\eqref{HHD},
as the terms involving the  conformal anomaly are removed,
\begin{subequations}\label{HHDp}
\begin{align}
[S_+^{(0)},\mathbf{H}^{(1)}] & = 0,
\label{S0H1}\\
[S_+^{(0)},\mathbf{H}^{(2)}] & =
\big[\mathbf{H}^{(1)},z_1+z_2\big]\left(\beta_0+\frac12 \mathbf{H}^{(1)}\right), 
\label{S0H2}
\\
[S_+^{(0)},\mathbf{H}^{(3)}] & =
\big[\mathbf{H}^{(1)},z_1+z_2\big]\left(\beta_1+ \frac12\mathbf{H}^{(2)}\right)+\big[\mathbf{H}^{(2)},z_1
+z_2\big]\left(\beta_0+ \frac12\mathbf{H}^{(1)}\right).
\end{align}
\end{subequations}
These equations are solved by
\begin{align}
\label{boldH-solve}
 \mathbf{H}^{(1)} & = \mathbf{H}^{(1)}_{\rm inv}\,,
\notag\\
  \mathbf{H}^{(2)} & = \mathbf{H}^{(2)}_{\rm inv} + \mathbb{T}^{(1)}\left(\beta_0+\frac12 \mathbf{H}_{\rm inv}^{(1)}\right),
\\
  \mathbf{H}^{(3)} & = \mathbf{H}^{(3)}_{\rm inv} + \mathbb{T}^{(1)} \left(\beta_1+\frac12 \mathbf{H}_{\rm inv}^{(2)}\right)
+\mathbb{T}^{(2)}_1 \left(\beta_0+\frac12 \mathbf{H}_{\rm inv}^{(1)}\right)^2
+\left(\mathbb{T}^{(2)}+\frac12 \big(\mathbb{T}^{(1)}\big)^2\right) \left(\beta_0+\frac12
\mathbf{H}_{\rm inv}^{(1)}\right)\,,
\notag
\end{align}
where $\mathbf{H}^{(k)}_{\rm inv}$ are (canonically) $SL(2)$-invariant operators with kernels that are functions of the
conformal ratio~\eqref{hinv} and the operators $\mathbb{T}^{(i)}$ commute with $S_-^{(0)}$ and $S_0^{(0)}$ and obey
the following equations:
\begin{align}\label{T-eq}
[S_+^{(0)},\mathbb{T}^{(1)}]   &= [\mathbf{H}_{\rm inv}^{(1)},z_1+z_2],
\notag\\
[S_+^{(0)},\mathbb{T}^{(2)}]   &= [\mathbf{H}_{\rm inv}^{(2)},z_1+z_2]\,,
\qquad\qquad
[S_+^{(0)},\mathbb{T}^{(2)}_1] =[\mathbb{T}^{(1)},z_1+z_2]\,.
\end{align}
Similar to the $\mathbb{X}$ kernels defined as the solutions to Eqs.~\eqref{X12def},
the $\mathbb{T}$ kernels are fixed by Eqs.~\eqref{T-eq} 
up to $SL(2)$ (canonically) invariant terms.
Explicit expressions are collected in App.~\ref{app:T}.

Note that the expressions for the perturbative expansion of the evolution kernel in Eq.~\eqref{boldH-solve}
can be assembled in the following single expression:
\begin{align}\label{H-N-I}
\bar\beta(a)+\frac12\mathbf{H}(a)=
\left\{\II-\frac12 \left(a \mathbb{T}^{(1)} +a^2 \left( \mathbb{T}^{(2)}+ \mathbb{T}^{(2)}_1 \mathbf{H}_{\rm inv}(a) \right)+O(a^3)\right)\right\}^{-1}\,
\biggl(\bar\beta(a)+\frac12 \mathbf{H}_{\rm inv}(a)\biggr).
\end{align}
Finally, adding the contributions from the rotation matrix $\mathrm{U} = \exp\{ a \mathbb{X}^{(1)}+ a^2 \mathbb{X}^{(2)}+\ldots\}$ we obtain the
following results for the first three orders of the evolution kernel in the $\overline{\text{MS}}$ scheme:
\begin{align}
 \mathbb{H}^{(1)} &= \mathbf{H}^{(1)} = \mathbf{H}^{(1)}_{\rm inv}\,,
\notag\\
\mathbb{H}^{(2)}& = \mathbf{H}^{(2)} +[\mathbf{H}^{(1)},\mathbb{X}^{(1)}]
 =\mathbf{H}^{(2)}_{\rm inv}+ \mathbb{T}^{(1)}\Big(\beta_0 +\frac12 \mathbf{H}_{\rm inv}^{(1)}\Big)+[\mathbf{H}_{\rm inv}^{(1)},\mathbb{X}^{(1)}]\,,
\notag\\
 \mathbb{H}^{(3)} & =  \mathbf{H}^{(3)} + [\mathbf{H}^{(2)},\mathbb{X}^{(1)}]  + [\mathbf{H}^{(1)},\mathbb{X}^{(2)}] + \frac12 \{\mathbf{H}^{(2)},(\mathbb{X}^{(1)})^2\}
\notag\\&=
\mathbf{H}^{(3)}_{\rm inv} + \mathbb{T}^{(1)} \left(\beta_1+\frac12 \mathbf{H}_{\rm inv}^{(2)}\right)
+\mathbb{T}^{(2)}_1 \left(\beta_0+\frac12 \mathbf{H}_{\rm inv}^{(1)}\right)^2
+\left(\mathbb{T}^{(2)}+\frac12 \big(\mathbb{T}^{(1)}\big)^2\right) \left(\beta_0+\frac12 \mathbf{H}_{\rm inv}^{(1)}\right)
\notag\\
&\quad +
[\mathbf{H}_{\rm inv}^{(2)}, \mathbb{X}^{(1)}]
+\frac12 \big[ \mathbb{T}^{(1)}\mathbf{H}_{\rm inv}^{(1)},\mathbb{X}^{(1)}\big] +\frac12 \big[\mathbf{H}_{\rm inv}^{(1)},\mathbb{X}^{(2,1)}\big]
\mathbf{H}^{(1)}_{\rm inv}
+[\mathbf{H}^{(1)}_{\rm inv}, \mathbb{X}_{\rm I}^{(2)}]
\notag\\
&\quad
+\beta_0\Big(\big[\mathbf{T}_{\rm inv}^{(1)},\mathbb{X}^{(1)}\big]+
\big[\mathbf{H}_{\rm inv}^{(1)},\mathbb{X}^{(2,1)}\big]\Big) +\frac12 \big[\big[\mathbf{H}_{\rm inv}^{(1)},\mathbb{X}^{(1)}\big],\mathbb{X}^{(1)}\big]
-\frac12 \big[\mathbf{H}_{\rm inv}^{(1)},\mathbb{X}^{(2,2)}\big],
\label{H-MSbar}
\end{align}
where all entries are known except for the $SL(2)$-invariant part of the three-loop kernel $\mathbf{H}^{(3)}_{\rm inv}$ that has yet to be determined.
Explicit expressions for the $\mathbb{X}$ and $\mathbb{T}$ kernels are given in App.~\ref{app:X} and App.~\ref{app:T}, respectively.%
\footnote{
The $\mathbb{X}^{(2)}$ kernels with an extra index, $\mathbb{X}_{\rm I}^{(2)}$, $\mathbb{X}^{(2,1)}$  and $\mathbb{X}^{(2,2)}$,
correspond to different contributions to $\mathbb{X}^{(2)}$ as described in App.~\ref{app:X}.}
The $SL(2)$-invariant kernels $\mathbf{H}^{(k)}_{\rm inv}$ can be written in the following general form
\begin{align}\label{Hinvk}
\mathbf{H}_{\rm inv}^{(k)} f (z_1,z_2) &=
\Gamma_{\rm cusp}^{(k)} \int_0^1d\alpha\frac{\bar\alpha}{\alpha} \Big(2f(z_1,z_2)-f(z_{12}^\alpha,z_2)-f(z_1,z_{21}^\alpha)\Big)
+ \chi^{(k)}_0 f (z_1,z_2)
\notag\\
&\quad +\int_0^1d\alpha \int_0^{\bar\alpha} d\beta \Big(\chi_{\rm inv}^{(k)}(\tau)+\chi_{\rm inv}^{\mathbb{P}(k)}(\tau) \mathbb{P}_{12}\Big) f(z_{12}^\alpha,z_{21}^\beta)\,.
\end{align}
Here $z_{12}^\alpha$ is defined in Eq.~\eqref{z12alpha}, $\tau=\alpha\beta/(\bar\alpha\bar\beta)$ and $\mathbb{P}_{12}$ is the
permutation operator, $\mathbb{P}_{12} f(z_1,z_2)= f(z_2,z_1)$.
$\Gamma_{\rm cusp}$ is the cusp anomalous dimension which is known to the required accuracy~\cite{Moch:2004pa}:
\begin{align}
 \Gamma_{\rm cusp}^{(1)} &= 4 C_F\,,
\notag\\
 \Gamma_{\rm cusp}^{(2)} &= 16 \left[C_A C_F \left(\frac{67}{36}-\frac{\pi^2}{12}\right)- \frac5{18} n_f C_F\right],
\notag\\
 \Gamma_{\rm cusp}^{(3)} &= 64 \biggl[
 C_{A}^2 C_{F}  \left( \frac{245}{96}- \frac{67 \pi^2}{216} + \frac{11 \pi^4}{720} + \frac{11}{24} \zeta_3 \right)
 +  C_{A} C_{F} n_f \frac12 \left( -\frac{209}{216} + \frac{5 \pi^2}{54} - \frac{7}{6}\zeta_{3} \right)
\notag\\&\quad
 + C^2_{F} n_f \frac12 \left( \zeta_3 -\frac{55}{48} \right) - \frac{1}{108} C_{F} n_f^2
\biggr].
\label{Gamma-cusp}
\end{align}
The LO expression \eqref{Honeloop} corresponds to
\begin{align}
 \chi_0^{(1)} = 2 C_F\,, && \chi_{\rm inv}^{(1)}(\tau) = -4 C_F\,, &&  \chi_{\rm inv}^{\mathbb{P}(1)}(\tau) = 0\,.
\label{chi-inv-1}
\end{align}
The two-loop constant term  $\chi_0^{(2)}$ and the functions $\chi_{\rm inv}^{(2)}(\tau)$, $\chi_{\rm
inv}^{\mathbb{P}(2)}(\tau)$ are given in App.~\ref{app:H2loop}. The three-loop expressions will be derived below.

Note that the expression for the two-loop kernel in Eq.~\eqref{H-MSbar} differs from the one derived in Ref.~\cite{Braun:2014vba}
where the non-invariant part is written in form of a single expression. The representation of the non-invariant
part of the two- and three-loop kernel as a product of simpler operators suggested here seems to be sufficient and probably
more convenient for most applications.

\section{Reciprocity relation}\label{sec:reciprocity}

The eigenvalues of  $\mathbb{H}^{(k)}$
\begin{align}
    \mathbb{H}^{(k)} (z_1-z_2)^N &=  \gamma^{(k)}(N) \,  (z_1-z_2)^N
\label{adim}
\end{align}
correspond to the flavor-nonsinglet anomalous dimensions in the $\overline{\text{MS}}$ scheme that are known to
three-loop accuracy~\cite{Moch:2004pa}. Note that our definition of the anomalous dimension $\gamma^{(k)}(N)$
differs from the one used in Ref.~\cite{Moch:2004pa} by an overall factor of two. In addition, in our work $N$
refers to the number of derivatives whereas in~\cite{Moch:2004pa} the anomalous dimensions are given as functions of
Lorentz spin of the operator. Thus
\begin{align}
    \gamma^{(k)}(N)\Big|_{\rm this~work} =  2\,\gamma^{(k)}(N+1)\Big|_{\text{Ref.~\cite{Moch:2004pa}}}.
\end{align}
One can show that the eigenvalues of the invariant kernels  $\mathbf{H}^{(3)}_{\rm inv}$
\begin{align}
   \mathbf{H}_{\rm inv}^{(k)} (z_1-z_2)^N =  \gamma_{\rm inv}^{(k)}(N) \,  (z_1-z_2)^N
\label{adim-inv}
\end{align}
with a ``natural'' choice of $\mathbb{T}$ operators specified in App.~\ref{app:T}
satisfy the following symmetry relation: Let $j_N=N+2$ (conformal spin); the asymptotic expansion of $\gamma_{\rm inv}^{(k)}(N)$ at large $j\to\infty$
only contains terms symmetric under the replacement $j_N\to 1-j_N$. Note that this symmetry does not hold for the anomalous
dimensions $\gamma^{(k)}(N)$ themselves.

The argument goes as follows.
As well known~\cite{Bukhvostov:1985rn}, conformal symmetry implies that the evolution kernels can be expressed in terms of the quadratic Casimir operator of the
symmetry group.
As a consequence it is natural to parameterize  the anomalous dimensions in the form
\begin{align}\label{gamma-f}
\gamma(N)=f\Big(N+2+\bar\beta(a)+\frac12\gamma(N)\Big)=f\Big(j_N+\bar\beta(a)+\frac12\gamma(N)\Big)\,.
\end{align}
It has been  shown~\cite{Basso:2006nk,Alday:2015eya,Alday:2015ewa} 
that the asymptotic expansion of the function $f(j)=a f^{(1)}(j)+a^2 f^{(2)}(j)+\ldots$ at large $j$ only contains terms that are symmetric
under reflection $j\to 1-j$. In all known examples the $f$-function also proves to be simpler than the anomalous
dimension itself. Expanding both sides of Eq.~\eqref{gamma-f} in a power
series in the coupling one obtains
\footnote{See also Ref.~\cite{Dokshitzer:2005bf} for a derivation in terms of the corresponding splitting functions in $x$-space.}
\begin{subequations}
\label{gamma-f-N}
\begin{align}
\label{f1}
f^{(1)}(j_N) & = \gamma^{(1)}(N)\,,
\\
\label{f2}
f^{(2)} (j_N) &= \gamma^{(2)}(N)-\frac{d}{dN} \left(\beta_0+\frac12 \gamma^{(1)}(N)\right)^2
             =\gamma^{(2)}(N)- \left(\beta_0+\frac12 \gamma^{(1)}(N)\right)\frac{d}{dN} f^{(1)}(j_N)\,,
\\
\label{f3}
f^{(3)}(j_N) &=\gamma^{(3)}(N) -\left(\beta_1+\frac12 \gamma^{(2)}(N)\right) \frac d{dN} f^{(1)}(j_N)
              - \frac12 \left(\beta_0+\frac12 \gamma^{(1)}(N)\right)^2 \frac{ d^2}{dN^2} f^{(1)}(j_N)
\notag\\
           &\quad
              -\left(\beta_0+\frac12 \gamma^{(1)}(N)\right) \frac d{dN} f^{(2)}(j_N)\,,
\end{align}
\end{subequations}
so that the values of $f^{(k)}(j_N)$ are related to the anomalous dimensions 
at the same order of perturbation theory up to
subtractions of certain lower-order terms.

Let us compare this expansion with the relations between the eigenvalues of $\mathbb{H}^{(k)}$ in Eq.~\eqref{adim}
and of the invariant kernel $\mathbf{H}_{\rm inv}^{(k)}$ in Eq.~\eqref{adim-inv}.
Using Eq.~\eqref{H-MSbar} and explicit expressions for the eigenvalues of the $\mathbb{T}$ kernels in Eq.~\eqref{adim-T}
one obtains (note that the commutator terms do not contribute to the spectrum)
\begin{align}
  \gamma^{(1)}(N) &=  \gamma_{\rm inv}^{(1)}(N)\,,
\notag\\
  \gamma^{(2)}(N) &=  \gamma_{\rm inv}^{(2)}(N) + \Big(\beta_0 +\frac12 \gamma^{(1)}(N) \Big)  \frac{d}{dN}\gamma_{\rm inv}^{(1)}(N)\,,
\notag\\
  \gamma^{(3)}(N) &=  \gamma_{\rm inv}^{(3)}(N) + \left(\beta_1+\frac12  \gamma_{\rm inv}^{(2)}(N)\right)  \frac{d}{dN}\gamma_{\rm inv}^{(1)}(N)
+ \frac12 \left(\beta_0+\frac12  \gamma_{\rm inv}^{(1)}(N)\right)^2 \frac{d^2}{dN^2}\gamma_{\rm inv}^{(1)}(N)
\notag\\&\quad
+ \left(\beta_0+\frac12  \gamma_{\rm inv}^{(1)}(N)\right) \biggl[\frac{d}{dN}\gamma_{\rm inv}^{(2)}(N)  +\frac12 \left(\frac{d}{dN}\gamma_{\rm inv}^{(1)}(N)\right)^2\biggr].
\end{align}
Comparing this expansion to the one in Eq.~\eqref{gamma-f-N} we see that
\begin{align}
    f^{(k)}(j_N) &= \gamma_{\rm inv}^{(k)}(N)\,.
\end{align}
In other words, the QCD anomalous dimension in the $\overline{\text{MS}}$ scheme can be written in terms of the
eigenvalues of the invariant kernel as
\begin{align}
\gamma(N)=\gamma_{\rm inv}\Big(N+2+\bar\beta(a)+\frac12\gamma(N)\Big),
\label{reciprocity2}
\end{align}
and the asymptotic expansion of $\gamma_{\rm inv}(N)$ at large $N$ only contains terms invariant under the reciprocity transformation
$j_N = N+2 \mapsto 1-j_N = -N-1$~\cite{Basso:2006nk,Alday:2015eya,Alday:2015ewa}.

This relation implies a certain condition for the choice of $\mathbb{T}$ kernels which appears to be natural
in the first few orders of perturbation theory, see App.~\ref{app:T}.

\section{Three-loop invariant kernel $\mathbf{H}^{(3)}_{\rm inv}$ }\label{sec:H3inv}

The three-loop invariant kernel $\mathbf{H}^{(3)}_{\rm inv}$ takes the form
\begin{align}\label{Hinv3}
\mathbf{H}_{\rm inv}^{(3)} f (z_1,z_2) &=
\Gamma_{\rm cusp}^{(3)} \int_0^1d\alpha\frac{\bar\alpha}{\alpha} \Big(2f(z_1,z_2)-f(z_{12}^\alpha,z_2)-f(z_1,z_{21}^\alpha)\Big)
+ \chi^{(3)}_0 f (z_1,z_2)
\notag\\
&\quad +\int_0^1d\alpha \int_0^{\bar\alpha} d\beta \Big(\chi_{\rm inv}^{(3)}(\tau)+\chi_{\rm inv}^{\mathbb{P}(3)}(\tau) \mathbb{P}_{12}\Big) f(z_{12}^\alpha,z_{21}^\beta)\,.
\end{align}
The three-loop cusp anomalous dimension $\Gamma_{\rm cusp}^{(3)}$ is known \eqref{Gamma-cusp} so that our task is to determine
the constant $\chi^{(3)}_0$ and two functions of one variable, $\chi_{\rm inv}^{(3)}(\tau)$ and $\chi_{\rm inv}^{\mathbb{P}(3)}(\tau)$.
This can be done by using the information on the spectrum of $\mathbf{H}_{\rm inv}^{(3)}$
\begin{align}
   \mathbf{H}_{\rm inv}^{(3)} (z_1-z_2)^N =  \gamma_{\rm inv}^{(3)}(N) \,  (z_1-z_2)^N\,,
\end{align}
where
\begin{align}
  \gamma_{\rm inv}^{(3)}(N) &= \gamma^{(3)}(N) -\left(\beta_1+\frac12 \gamma^{(2)}(N)\right) \frac d{dN} \gamma^{(1)}(N)
              -\left(\beta_0+\frac12 \gamma^{(1)}(N)\right) \frac d{dN} \gamma^{(2)}(N)
\notag\\&\quad
              +\frac12 \left(\beta_0+\frac12 \gamma^{(1)}(N)\right) \left(\frac{d}{dN} \gamma^{(1)}(N)\right)^2
\label{gamma3-inv}
\end{align}
can easily be calculated from the known one-, two-, and three-loop flavor-nonsinglet anomalous dimensions~\cite{Moch:2004pa}.

The constant term $\chi^{(3)}_0$ in the invariant kernel~\eqref{Hinv3} corresponds to the constant term in the  large-$N$ asymptotic of the anomalous dimension
\begin{align}
    \gamma_{\rm inv}^{(3)}(N) &=  2\,\Gamma^{(3)}_{\rm cusp}\Big[\psi(N+2)-\psi(2)\Big] + \chi^{(3)}_0 +O(1/N^2)\,
\end{align}
and is straightforward to obtain. Using the expressions from Ref.~\cite{Moch:2004pa} we find
\begin{align}
  \chi^{(3)}_0 &= C_F^3\Big[\dfrac{3176}{9} \zeta _3-320 \zeta_5+\dfrac{1672 \pi ^4}{135}
-\dfrac{23954 \pi^2}{81}+\dfrac{13454}{9}\Big]
\notag\\&\quad
              + C_F^2 n_f\Big[-\dfrac{752}{9} \zeta _3-\dfrac{128 \pi ^4}{135}
+\dfrac{3452 \pi^2}{81}-\dfrac{6242}{27}\Big]
              + C_F n^2_f\Big[\dfrac{32}{9} \zeta _3-\dfrac{80 \pi ^2}{81}+\dfrac{70}{27}\Big]
\notag\\&\quad
              + \dfrac{C^2_F}{N_c}\Big[-\dfrac{16}{3} \pi ^2 \zeta _3+\dfrac{9464}{9} \zeta _3-560 \zeta _5+\dfrac{322\pi ^4}{27}-\dfrac{27158 \pi ^2}{81}+\dfrac{28789}{18}\Big]
\notag\\&\quad
              + \dfrac{C_F n_f}{N_c}\Big[-\dfrac{1072}{9} \zeta _3-\dfrac{2 \pi ^4}{45}
+\dfrac{1816 \pi^2}{81}-\dfrac{2752}{27}\Big]
\notag\\&\quad
              + \dfrac{C_F}{N^2_c}\Big[\dfrac{3632}{9} \zeta _3-80 \zeta _5+\dfrac{31 \pi ^4}{15}-\dfrac{7712 \pi^2}{81}+\dfrac{7537}{18}\Big].
\label{chi3-0}
\end{align}
The calculation of the functions $\chi_{\rm inv}^{(3)}(\tau)$ and $\chi_{\rm inv}^{\mathbb{P}(3)}(\tau)$ is much more involved.
As usual one has to consider even and odd values of $N$ separately. We write
\begin{align}
  \gamma_{\rm inv}^{(3)}(N) =  \gamma_{\rm inv }^{(3+)}(N) + (-1)^N  \gamma_{\rm inv }^{(3-)}(N)\,,
\end{align}
so that the combinations $\gamma_{\rm inv }^{(3+)}(N) \pm \gamma_{\rm inv }^{(3-)}(N) $ correspond to the eigenvalues of the
invariant kernel for even (odd) $N$.
Using the representation in \eqref{Hinv3} one obtains
\begin{subequations}
\begin{align}
 \gamma_{\rm inv }^{(3+)}(N) & = 2 \Gamma^{(3)}_{\rm cusp}\big[\psi(N+2)-\psi(2)\big] + \chi_0^{(3)}
     +\int_0^1 d\alpha \int_0^{\bar\alpha} d\beta\, \chi^{(3)}_{\rm inv}(\tau)\, (1-\alpha-\beta)^{N}\,,
\label{ffplus} \\
\gamma_{\rm inv }^{(3-)}(N)  & =\int_0^1 d\alpha \int_0^{\bar\alpha} d\beta \,\chi_{\rm inv}^{\mathbb{P}(3)}(\tau)\, (1-\alpha-\beta)^{N}.
\label{ffminus}
\end{align}
\end{subequations}
These relations can be inverted to express the kernels as functions of the anomalous dimensions~\cite{Braun:2014vba}
\begin{align}\label{Legendre}
\chi^{(3)}_{\rm inv}(\tau) &= \frac{1}{2\pi i}\int_{c-i\infty}^{c+i\infty} dN \,(2N+3)\,
 \Delta \gamma_{\rm inv }^{(3+)}(N)\,
P_{N+1}\left(\frac{1+\tau}{1-\tau}\right),
\notag\\
\chi_{\rm inv}^{\mathbb{P}(3)}(\tau) &= \frac{1}{2\pi i}\int_{c-i\infty}^{c+i\infty} dN \,(2N+3)\,
\gamma_{\rm inv }^{(3-)}(N)\, P_{N+1}\left(\frac{1+\tau}{1-\tau}\right)\,,
\end{align}
%
%
%
where
\begin{align}
 \Delta \gamma_{\rm inv }^{(3+)}(N) =
    \gamma_{\rm inv }^{(3+)}(N) - 2 \Gamma^{(3)}_{\rm cusp}\big[\psi(N+2)-\psi(2)\big] -  \chi_0^{(3)}\,,
\end{align}
and $P_{N+1}$ is the Legendre function.  All singularities of the anomalous dimensions have to lie to the left of the
integration contour.

The algebraic structure of the three-loop anomalous dimension $\gamma_{\rm inv }^{(3)}(N)$~\cite{Moch:2004pa} is, unfortunately, too complicated to do
the integrals in~\eqref{Legendre} analytically. We, therefore, adopted the following
strategy: We will provide analytic expressions for the terms that correspond to
\begin{enumerate}
\item the leading contributions  $\sim\dfrac{(\psi(j)-\psi(2))^k}{j(j-1)}$, $j=N+2$, to the large-$N$ expansion of
    the anomalous dimensions,
\item the contribution of the leading singularity (pole at $N=-1$) in the complex plane.
\end{enumerate}
The remainder will be parameterized by a sufficiently simple function with a few fit parameters.

\subsection{Splitting functions}\label{sec:split}

It turns out to be advantageous to use the representation for the anomalous dimensions in terms
of the splitting functions. The three-loop splitting functions are available from Ref.~\cite{Moch:2004pa} and involve harmonic polylogarithms
(HPL) up to weight five, see Ref.~\cite{Remiddi:1999ew}.
 Using this result and Eq.~\eqref{gamma3-inv} it is straightforward to calculate the splitting
functions for the invariant kernels such that
\begin{align}
  \gamma_{\rm inv}^{(3\pm)}(N) = -\int_0^1 dx\, x^N  H_{\rm inv}^{(3\pm)}(x)\,.
\label{def:split}
\end{align}
The ``plus'' function can be written as
\begin{align}
 H_{\rm inv}^{(3+)}(x) &= 2 \Gamma^{(3)}_{\rm cusp} \,\frac{x}{(1-x)_+} - \big(\chi^{(3)}_0 - 2 \Gamma^{(3)}_{\rm cusp}\big)\delta(1-x) +  \Delta H_{\rm inv}^{(3+)}(x)\,,
\end{align}
where the first two terms are related to the logarithmic and the constant contributions in the  large-$N$ expansion of the anomalous dimension,
cf.~\eqref{ffplus} and for $H_{\rm inv}^{(3-)}(x)$ there are no such terms, cf.~\eqref{ffminus}.
The symmetry property under $j_N\mapsto 1-j_N$ for the eigenvalues of the invariant kernel
is equivalent to the Gribov-Lipatov reciprocity relation for the corresponding splitting functions
\begin{align}\label{rec-rel}
\Delta H_{\rm inv}^{(3\pm)}(x) = - x \Delta H_{\rm inv}^{(3\pm)}(1/x)\,.
\end{align}
We want to find a parametrization for the splitting functions consistent with the reciprocity relations \eqref{rec-rel} and
separating the leading contributions at $x\to 0$ and $x\to 1$.

To this end, let us define the set of functions $\phi_k(x)$ such that
\begin{align}
\int_0^1 dx\,x^{j-2}\, \phi_k(x) = \left(\frac1{j(j-1)}\right)^{k+1}\,.
\end{align}
They can be constructed recursively,
\begin{align}\label{f-k}
\phi_0(x)=1-x\,, \qquad \phi_k(x)=\int_x^1 \frac{d\xi}{\xi} \phi_{k-1}(\xi) \phi_0(x/\xi)\,,
\end{align}
so that
\begin{align}
   \phi_1(x) &= -2\bar x - (1+x) \ln x\,,
\notag\\
   \phi_2(x) &= 2\bar x^2  + \bar x (3+x) \ln x + (1+x) \ln^2 x\,,
\end{align}
where here and below
\begin{align}
   \bar x = 1-x\,.
\end{align}
We write the splitting functions as
\begin{align}\label{Hx}
\Delta H_{\rm inv}^{(3+)}(x) &= \sum_{k=1}^{4} B^{(3+)}_k \phi_k(x) + \bar x\,C^{(3+)}_0 + \bar x\, C^{(3+)}_1 \ln\left(\frac{x}{\bar x^2}\right) +\delta  H_{\rm inv}^{(3+)}(x)\,,
\notag\\
       H_{\rm inv}^{(3-)}(x) &= \sum_{k=1}^{4} B^{(3-)}_k \phi_k(x) + \bar x\, C^{(3-)}_0 + \delta  H_{\rm inv}^{(3-)}(x)\,,
\end{align}
where the addenda, $\delta  H_{\rm inv}^{(3+)}(x)$ and $\delta  H_{\rm inv}^{(3-)}(x)$, do not include, by construction, terms $\ln^k x$, $k\ge 1$ (for $x\to 0$)
and $\bar x\,\ln^k\bar x $, $k\ge 0$ (for $x\to1$).
The maximum powers of the logarithms are found by inspection of the known analytic expression.
Thus, with normalization constants $H_0^\pm$
\begin{align} \label{dHconstraint}
\delta H_{\rm inv}^{(3\pm)}(x)\underset{x\to 1}{=} \mathcal{O}(\bar x^3)\,, &&  
\delta H_{\rm inv}^{(3\pm)}(x)\underset{x\to 0}{=} 
H_0^\pm + \mathcal{O}(x)\,,
\end{align}
or, equivalently, in moment space,
\begin{align}\label{gjH}
\delta\gamma_{\rm inv}^{(3\pm)}(N)=-\int_0^1 dx \, x^{N}\, \delta H_{\rm inv}^{(3\pm)}(x)\,,
\end{align}
vanishes as $1/N^4$ at large $N$ and its only possible singularity at $N=-1$ is a simple pole.

The constants $B^{(3\pm)}_k$,  $C^{(3\pm)}_k$ are collected in Table~\ref{tab:resultcoeff} where, for completeness, we also repeat the
expressions for $\chi^{(3)}_0$ \eqref{chi3-0} and $\Gamma^{(3)}_{\rm cusp}$ \eqref{Gamma-cusp}.
In all cases we show the coefficients for the following color decomposition:
\begin{align}
  F &= C_F^3 F_{\langle 1\rangle} + C_F^2 n_f F_{\langle 2\rangle} +  C_Fn^2_f F_{\langle 3\rangle}
       + \frac{C_F^2}{N_c}  F_{\langle 4\rangle} +   \frac{C_F n_f}{N_c} F_{\langle 5\rangle}
       + \frac{C_F}{N^2_c} F_{\langle 6\rangle}\,,
\label{colorstructures}
\end{align}
where $F = A^{(3\pm)}_k,\, B^{(3\pm)}_k,\, \chi^{(3)}_0,\, \Gamma^{(3)}_{\rm cusp}$.

\begin{table}[t]
\renewcommand{\arraystretch}{2.0}
\begin{subtable}[h]{\linewidth}
\centering
\scalebox{0.85}{
\begin{tabular}{|C|C|C|}
\hline\hline
  & \chi_0^{(3)} & \Gamma^{(3)}_{\rm cusp} \\[1mm] \hline
 C_F^3 &
\dfrac{3176}{9} \zeta _3 - 320 \zeta _5+\dfrac{1672 \pi ^4}{135}-\dfrac{23954 \pi
   ^2}{81}+\dfrac{13454}{9}
       & \dfrac{352}{3}\zeta_3+\dfrac{1960}{3}-\dfrac{2144 \pi ^2}{27}+\dfrac{176 \pi ^4}{45} \\[1mm] \hline
 C_F^2 n_f & -\dfrac{752 }{9}\zeta _3-\dfrac{128 \pi ^4}{135}+\dfrac{3452 \pi
   ^2}{81}-\dfrac{6242}{27}
         & -\dfrac{128}{3}\zeta_3-\dfrac{2662}{27}+\dfrac{160 \pi ^2}{27}
\\ \hline
 C_F n^2_f & \dfrac{32}{9} \zeta _3-\dfrac{80 \pi ^2}{81}+\dfrac{70}{27}
             & -\dfrac{16}{27}
\\[1mm] \hline
 \dfrac{C_F^2}{N_c} & -\dfrac{16 \pi^2}{3}\zeta_3 +\dfrac{9464}{9} \zeta _3-560 \zeta _5
+\dfrac{322\pi ^4}{27}-\dfrac{27158 \pi ^2}{81}+\dfrac{28789}{18}
                    & \dfrac{352}{3}\zeta_3+\dfrac{1960}{3}-\dfrac{2144 \pi ^2}{27}+\dfrac{176 \pi ^4}{45}
\\[1mm] \hline
 \dfrac{C_F n_f}{N_c} & -\dfrac{1072}{9} \zeta _3-\dfrac{2 \pi ^4}{45}+\dfrac{1816 \pi^2}{81}-\dfrac{2752}{27}
                      & -\dfrac{112}{3}\zeta_3-\dfrac{836}{27}+\dfrac{80 \pi ^2}{27}
\\[1mm] \hline
 \dfrac{C_F}{N_c^2} & \dfrac{3632}{9} \zeta _3-80 \zeta _5+\dfrac{31 \pi ^4}{15}-\dfrac{7712 \pi^2}{81}+\dfrac{7537}{18}
                    & \dfrac{88}{3}\zeta_3+\dfrac{490}{3}-\dfrac{536 \pi ^2}{27}+\dfrac{44 \pi ^4}{45}
   \\[1mm] \hline\hline
\end{tabular}
             }
\end{subtable}
\vskip0.5cm
\begin{subtable}[h]{\linewidth}
\centering
\scalebox{0.83}{
\begin{tabular}{|C|C|C|C|C|C|C|C|}
\hline\hline
           & B^{(3+)}_1 & B^{(3+)}_2 & \!B^{(3+)}_3\!\! & \!\!B^{(3+)}_4\!\! & C^{(3+)}_0& C^{(3+)}_1
\\[1mm] \hline
 \!C_F^3\! & -\dfrac{746}{9}-\dfrac{40 \pi ^2}{9} & 20-\dfrac{16 \pi ^2}{3} & 0 & 0 & \dfrac{352 }{3}\zeta_3+\dfrac{70768}{27}-\dfrac{3488 \pi ^2}{27}+\dfrac{176 \pi ^4}{45} & -\dfrac{184}{3}
\\[1mm] \hline
 \!C_F^2n_f\! & \dfrac{28}{9}-\dfrac{16 \pi ^2}{9} & 0 & 0 & 0 & -\dfrac{128}{3}\zeta_3-\dfrac{11966}{27}+\dfrac{256 \pi ^2}{27} & -\dfrac{16}{3}
\\[1mm] \hline
 \!C_Fn^2_f\! & 0 & 0 & 0 & 0 & \dfrac{64}{9} & 0
\\[1mm] \hline
 \!\dfrac{C_F^2}{N_c}\!\! & \!-176 \zeta_3+\dfrac{1886}{3}-\dfrac{52 \pi ^2}{9} & \!\dfrac{3632}{9}-\dfrac{16 \pi ^2}{3} & -\dfrac{520}{3} & -64 &
 \dfrac{352 }{3}\zeta_3+\dfrac{74428}{27}-\dfrac{3488 \pi ^2}{27}+\dfrac{176 \pi^4}{45} & \dfrac{16 \pi ^2}{3}-\dfrac{376}{3}
\\[1mm] \hline
\!\dfrac{C_F n_f}{N_c}\!\! & -\dfrac{512}{9}-\dfrac{8 \pi ^2}{9} & -\dfrac{400}{9} & -\dfrac{16}{3} & 0 & -\dfrac{112 }{3}\zeta_3-\dfrac{5932}{27}+\dfrac{128 \pi ^2}{27} & -\dfrac{8}{3}
\\[1mm] \hline
 \!\dfrac{C_F}{N^2_c}\!& \!-112 \zeta_3+\dfrac{3272}{9}-\dfrac{76 \pi ^2}{9} & \!\dfrac{1816}{9}-\dfrac{16 \pi ^2}{3} & -\dfrac{176}{3} & -24 &
\dfrac{88 }{3}\zeta_3+\dfrac{17902}{27}-\dfrac{836 \pi ^2}{27}+\dfrac{44 \pi^4}{45} & \dfrac{8 \pi ^2}{3}-\dfrac{196}{3}
\\[1mm] \hline\hline
\end{tabular}
}
\end{subtable}
\vskip0.5cm
\begin{subtable}[h]{\linewidth}
\centering
\scalebox{0.85}{
\begin{tabular}{|C|C|C|C|C|C|}
\hline\hline
           & B^{(3-)}_1 & B^{(3-)}_2 & B^{(3-)}_3 & B^{(3-)}_4 & C^{(3-)}_0
\\[1mm] \hline
\dfrac{C_F^2}{N_c}  & -128 \zeta_3+\dfrac{7288}{9}+\dfrac{52 \pi ^2}{9} & \dfrac{3632}{9}+\dfrac{32 \pi ^2}{3} & -\dfrac{520}{3} & -64 & \dfrac{88}{3}
\\[1mm] \hline
\dfrac{C_F n_f}{N_c}& \dfrac{16 \pi ^2}{9}-\dfrac{488}{9} & -\dfrac{400}{9} & -\dfrac{16}{3} & 0 & -\dfrac{8}{3}
\\[1mm] \hline
 \dfrac{C_F}{N^2_c} & -112 \zeta_3+\dfrac{3860}{9}+\dfrac{164 \pi ^2}{9} & \dfrac{1816}{9}+\dfrac{44 \pi ^2}{3} & -\dfrac{176}{3} & -24 & \dfrac{44}{3}
\\[1mm] \hline\hline
\end{tabular}
             }
\end{subtable}
\caption{Cusp anomalous dimension $\Gamma^{(3)}_{\rm cusp}$ \eqref{Gamma-cusp}, constant term  $\chi^{(3)}_0$ \eqref{chi3-0}
 and the coefficients $B^{(3\pm)}_k$,  $C^{(3\pm)}_k$ \eqref{Hx} in the splitting function representation of the
invariant kernel.}
\label{tab:resultcoeff}
\renewcommand{\arraystretch}{1.0}
\end{table}

\begin{figure}[t]
\scalebox{0.49}{\includegraphics[width=0.99\linewidth]{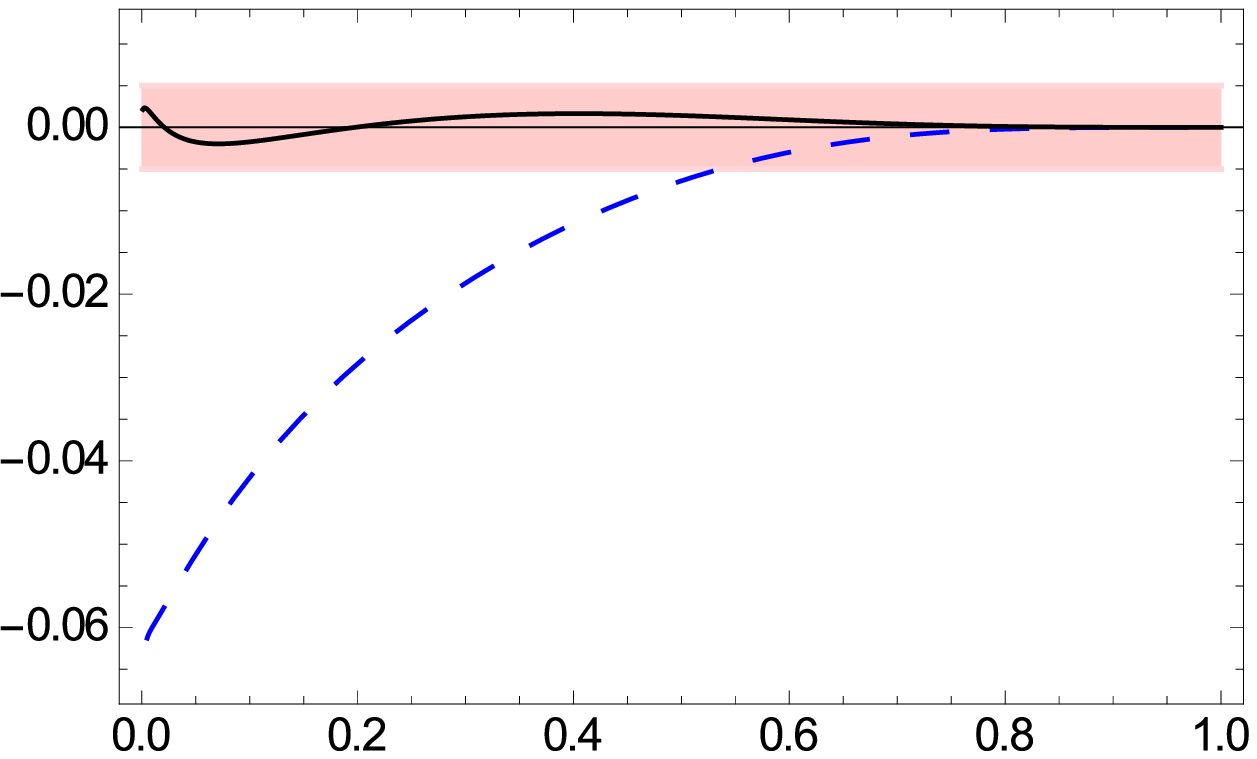}}~~
\scalebox{0.49}{\includegraphics[width=0.99\linewidth]{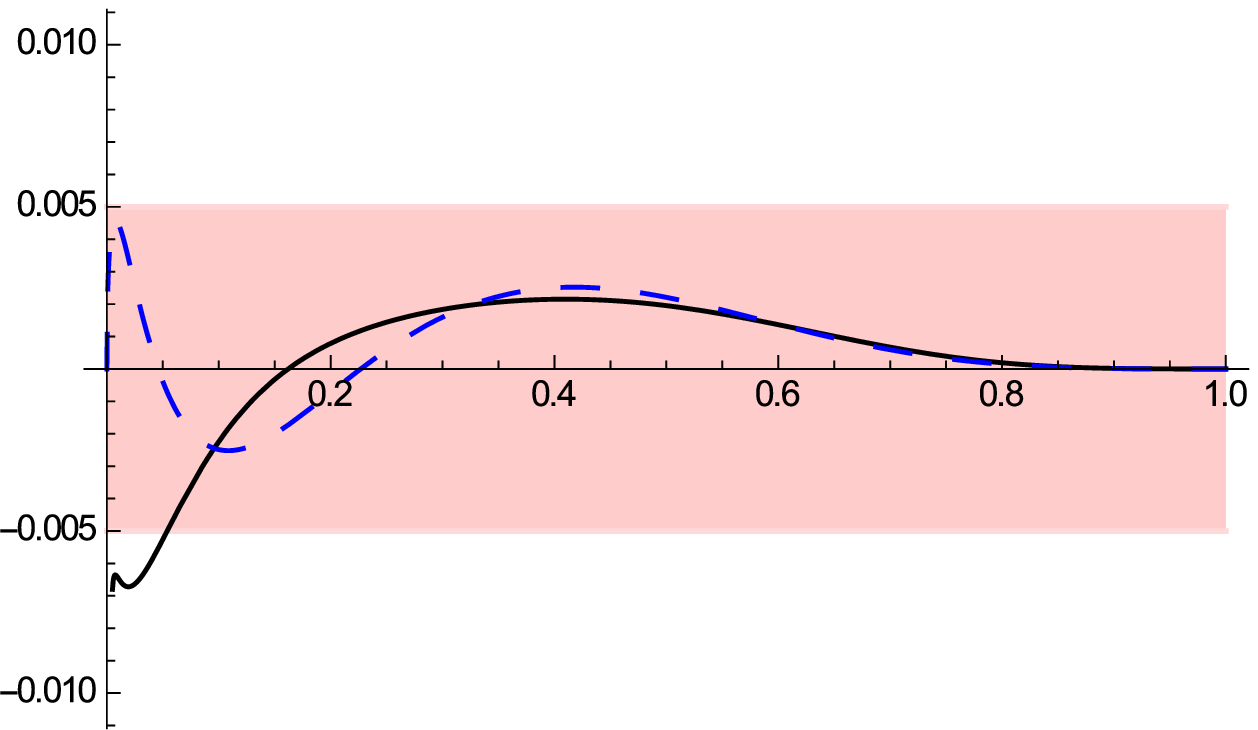}}
\caption{\small Left panel shows
the ratio $\delta H^{(3+)}_{\rm inv}(x)/H^{(3+)}_{\rm inv}(x)$ (dashed curve) for exact splitting functions and
the error in using the approximation \eqref{ansatz2},  $(\delta H^{(3+)}_{\rm inv}|_{\rm fit}- \delta H^{(3+)}_{\rm inv}|_{\rm exact}) /H^{(3+)}_{\rm inv}$
(solid curve) for $n_f=4$.  The shaded area indicates an error band of $0.5 \%$.
The similarly defined approximation error for the combinations  $H^{(+)} + H^{(-)}$ (dashes) and $H^{(+)} - H^{(-)}$ (solid) which
give rise to moments with odd and even $N$, respectively, is shown on the right panel.}
\label{fig:Error}
\end{figure}

The remaining terms  $\delta H^{(3\pm)}_{\rm inv}(x)$ contain the whole algebraic complexity of the full result but
numerically they are rather small. For illustration we show the ratio $\delta H^{(3+)}_{\rm inv}(x)/H^{(3+)}_{\rm inv}(x)$
for $N_c=3$ and $n_f=4$ in  Fig.~\ref{fig:Error} (dashed blue curve on the left panel). One sees that
$\delta H^{(3+)}_{\rm inv}(x)$ contributes at most $6\%$ to the full splitting function in the whole range $0<x<1$
so that for all practical purposes it can be approximated by a simple expression with a few parameters.

Due to the reciprocity property \eqref{rec-rel} the functions $\delta H^{(3\pm)}_{\rm inv}(x)$ can be parameterized in the form
\begin{align}
 \delta H^{(3\pm)}_{\rm inv}(x)= \bar x \, h_{\pm} (x/\bar x^2)\,.
\label{ansatz1}
\end{align}
We choose the following ansatz
\begin{align}
 h_\pm(t) =  H^\pm_0 \frac{a_\pm}{t+a_\pm} \left( 1 + \frac{b_\pm \,t}{t+a_\pm}\right) \,,
\label{ansatz2}
\end{align}
where $a_\pm$ and $b_\pm $ are fit parameters and the normalization constants $H_0^\pm$ are determined
analytically from the condition $\delta H_{\rm inv}^{(3\pm)}(x)\underset{x\to 0}{=} 
H_0^\pm + \mathcal{O}(x)$~\eqref{dHconstraint}.
The fitted values of the parameters $a_\pm$ and $b_\pm$ for the different color structures can be found
in Table \ref{tab:fitcoeff}. With this simple parametrization we reduce the deviation from the exact splitting
functions to less than 0.5\%, see~Fig. \ref{fig:Error}.

\begin{table}[t]
\renewcommand{\arraystretch}{2.0}
\begin{subtable}[h]{\linewidth}
\centering
\scalebox{0.85}{
\begin{tabular}{|C|C|C|C|}
\hline\hline
  & H_0^+ & a_+ & b_+ \\[1mm] \hline
  C_F^3 & \dfrac{272}{3}-\dfrac{28 \pi ^2}{9}-\dfrac{64 \pi ^4}{45} & 0.2263 & 0 \\[1mm] \hline
  C_F^2 n_f & \dfrac{32}{3}-\dfrac{16 \pi ^2}{9} & 0.5340 & 0 \\[1mm] \hline
 \dfrac{C_F^2}{N_c}& -\dfrac{368 \zeta_3}{3}-\dfrac{992}{9}+\dfrac{176 \pi ^2}{9}+\dfrac{4 \pi ^4}{9} & 0.05174 & 4.116 \\[1mm] \hline
 \dfrac{C_F n_f}{N_c}& -\dfrac{32 \zeta_3}{3}+\dfrac{256}{9}-\dfrac{8 \pi ^2}{9} & 0.09626 & -1.526 \\[1mm] \hline
 \dfrac{C_F^2}{N^2_c}& -\dfrac{328 \zeta_3}{3}-\dfrac{736}{9}+\dfrac{140 \pi ^2}{9}+\dfrac{8 \pi ^4}{5} & 0.06595 & 0
\\[1mm] \hline\hline
\end{tabular}
              }
\end{subtable}
\vskip0.5cm
\begin{subtable}[h]{\linewidth}
\centering
\scalebox{0.85}{
\begin{tabular}{|C|C|C|C|}
\hline\hline
  & H_0^- & a_- & b_- \\[1mm] \hline
  \dfrac{C_F^2}{N_c}  & 128 \zeta_3-24-\dfrac{2200 \pi ^2}{27}+\dfrac{28 \pi ^4}{9} & 0.4040 & -0.7986 \\[1mm] \hline
  \dfrac{C_F n_f}{N_c}& \dfrac{176 \pi ^2}{27}-16 \zeta_3 & 0.1252 & 0 \\[1mm] \hline
  \dfrac{C_F^2}{N^2_c} & 64 \zeta_3-24-\dfrac{1208 \pi ^2}{27}+\dfrac{7 \pi ^4}{9} & 0.2206 & -1.077
\\[1mm] \hline\hline
\end{tabular}
             }
\end{subtable}
\caption{Values of the parameters in the ansatz for $\delta H^{(3\pm)}_{\rm inv}(x)$~\eqref{ansatz1}, \eqref{ansatz2}.}
\label{tab:fitcoeff}
\renewcommand{\arraystretch}{1.0}
\end{table}
%

\subsection{Mellin transformation}

The following Mellin representation of the kernels $\chi_{\rm inv}(\tau)$
and $\chi_{\rm inv}^{\mathbb{P}}(\tau)$ proves to be useful in order to restore them from the splitting functions
and allows one to write all terms in the form that automatically respects the reciprocity relation:
\begin{subequations}
\label{mellin}
\begin{align}\label{mellin-transform-h}
\chi(\tau) & = \frac{ 1 }{2\pi i}\int_{c-i\infty}^{c+i\infty}\,d\rho\, \widetilde \chi(\rho) \left({\bar\tau}/{\tau}\right)^{-\rho}\,,
\\
\label{mellin-inverse-h}
 \widetilde \chi(\rho) & = \int_0^1 \frac{d\tau}{\tau\bar \tau} \left(\bar\tau/\tau\right)^\rho \, \chi(\tau)\,.
\end{align}
\end{subequations}
The integration contour in the first integral, \eqref{mellin-transform-h}, must be chosen in the analyticity strip of  the second
integral, \eqref{mellin-inverse-h}, (the strip where integral converges). Making use of this
representation  one obtains
\begin{subequations}
\label{mellin1}
\begin{align}
\Delta \gamma_{\rm inv }^{(3+)}(N) &
= \frac{1}{2\pi i}\int_{c-i\infty}^{c+i\infty}\,d\rho\, \widetilde \chi^{(3)}_{\rm inv}(\rho)  \Gamma^2(1+\rho)\frac{\Gamma(j_N-1-\rho)}{\Gamma(j_N+1+\rho)}
\,,
\\
\gamma_{\rm inv }^{(3-)}(N)  &
= \frac{1}{2\pi i}\int_{c-i\infty}^{c+i\infty}\,d\rho\, \widetilde \chi^{\mathbb{P}(3)}_{\rm inv}(\rho) \Gamma^2(1+\rho)\frac{\Gamma(j_N-1-\rho)}{\Gamma(j_N+1+\rho)}.
\end{align}
\end{subequations}
Since for large $j_N$ the ratio $\Gamma(j_N-1-\rho)/{\Gamma(j_N+1+\rho)}\sim j_N^{-2(1+\rho)}$, the asymptotic expansion
of the integrals in Eqs.~\eqref{mellin1} at $j_N\to \infty$ can be obtained by moving the integration contour to the right and picking up
the corresponding residues. It is easy to check that if the only singularities
of the Mellin-transformed kernels
$\widetilde \chi(\rho)$ in the right half-plane are poles (of arbitrary order) at real integer values of $\rho$,
then a generic term of the asymptotic expansion of the anomalous dimensions has the form
$$
  \frac{(\psi(j_N)-\psi(1))^m}{(j_N(j_N-1))^k}, \qquad k>0, m\geq 0\,,
$$
which is required by the reciprocity symmetry under the $j_N\mapsto 1-j_N$ transformation~\cite{Basso:2006nk,Alday:2015eya,Alday:2015ewa}.
Under the same condition ($\widetilde \chi(\rho)$ only has poles at integer $\rho$ values in the right half-plane),
the kernels $\chi(\tau)$ at small values of the conformal ratio have the expansion $$\chi(\tau)=\sum_{km}c_{km} \tau^k \ln^m\tau \,.$$

The Mellin transform of the one-loop kernel is very simple:
\begin{align}
  \chi_{\rm inv}^{(1)} = - 4 C_F \qquad\leftrightarrow\qquad    \widetilde \chi_{\rm inv}^{(1)} = - 4 C_F \Big[ 2\pi i \delta(\rho)\Big]\,.
  \end{align}
More examples are collected in Table~\ref{Tab:Examples}, see App.~\ref{app:H2loop}. One finds that the Mellin space
kernels have a rather simple form, while the corresponding anomalous dimensions can be quite involved. Also the
two-loop evolution kernels in Mellin space are given by rather compact expressions, see Eq.~\eqref{mellin-twoloop}.

If the anomalous dimensions are written in terms of the splitting functions, Eq.~\eqref{def:split},
\begin{align}
  \gamma_{\rm inv}^{(k\pm)}(N) = -\int_0^1 dx\, x^N  H_{\rm inv}^{(k\pm)}(x)\,,
\end{align}
the corresponding Mellin-transformed invariant kernels can be calculated as
\begin{align}\label{delta-rho}
 \widetilde{\chi}_{\rm inv}^{(k)}(\rho)
&=
-\frac{\Gamma(2\rho+2)}{\Gamma^2(1+\rho)}\int_0^1dx \, H_{\rm inv}^{(k+)}(x)\, \frac{1}{x\bar x} \left(\frac{1+x}{1-x}\right)
 \left(\frac x{\bar x^2}\right)^\rho\,
\end{align}
and similar for $ \widetilde{\chi}_{\rm inv}^{\mathbb{P}(k)}(\rho)$, with the replacement $H_{\rm inv}^{(k+)}(x)\to H_{\rm inv}^{(k-)}(x)$.
The kernels in $\tau$ space can finally be obtained by the inverse Mellin transformation~\eqref{mellin-transform-h}.
We found this two-step approach to be the most effective for the three-loop case.

For the remainder function $\delta H^{(3\pm)}_{\rm inv}(x)$ written in the form \eqref{ansatz1} one obtains in Mellin space
\begin{align}
\delta\widetilde{\chi}^{(3\pm)}_{\rm inv}(\rho)=
-\frac{\Gamma(2\rho+2)}{\Gamma^2(1+\rho)}\int_0^\infty dt\,  h_\pm(t) \, t^{\rho-1}\,,
\end{align}
and for the simple ansatz in Eq.~\eqref{ansatz2}
\begin{align}
\delta\widetilde{\chi}^{(3\pm)}_{\rm inv}(\rho) &=
-\frac{\Gamma(2\rho+2)}{\Gamma^2(1+\rho)} \frac{\pi}{\sin (\pi  \rho ) }
  H^\pm_0(1 + b_\pm \rho)a_\pm^{\rho } \,.
\end{align}
Using Eq.~\eqref{mellin-transform-h} we finally obtain the following
expression for the corresponding contribution to the invariant kernel
\begin{align}
\delta {\chi}^{(3\pm)}_{\rm inv}(\tau)
=
\frac{H^\pm_0}{\left(1+4a_\pm \tau/\bar \tau\right)^{5/2}}
\left[1 +a_\pm \frac{\tau}{\bar \tau}  (4-6\,b_\pm) \right] - H^\pm_0 \,.
\label{deltachi-tau}
\end{align}
We also need the expressions for the functions $\phi_k$~\eqref{f-k} in $\rho$- and $\tau$-space defined
as
\begin{align}
\widetilde{\phi}_k(\rho)
&\equiv
-\frac{\Gamma(2\rho+2)}{\Gamma^2(1+\rho)}\int_0^1dx \, \phi_k(x) \, \frac{1}{x\bar x} \left(\frac{1+x}{1-x}\right)
 \left(\frac x{\bar x^2}\right)^\rho
\equiv \int_0^1 \frac{d\tau}{\tau\bar \tau} \left(\bar\tau/\tau\right)^\rho \, \varphi_k(\tau)\,.
\end{align}
One obtains
\begin{align}
\widetilde{\phi}_1(\rho)&=-{\pi}/(\rho\sin\pi\rho),
\notag\\
\widetilde{\phi}_2(\rho)&=
\widetilde{\phi}_1(\rho)\, \frac{\bar\rho}{\rho},
\notag\\
\widetilde{\phi}_3(\rho)&=
\widetilde{\phi}_1(\rho)\,\Big( -2\,\frac{\bar\rho}{\rho}+\psi'(\rho)-\frac{\pi^2}{6}\Big),
\notag\\
\widetilde{\phi}_4(\rho)&=
\widetilde{\phi}_1(\rho)\,\Big( 5\,\frac{\bar\rho}{\rho}+\frac{(1-3\rho)}{\rho} \Big[\psi'(\rho)-\frac{\pi^2}{6}\Big]\Big),
\end{align}
and
\begin{align}
\varphi_1(\tau) & = \ln\bar\tau=H_1\left(-\frac{\tau}{\bar \tau}\right)\,,
\notag\\
\varphi_2(\tau) & = -\varphi_1(\tau) + H_{01}\left(-\frac{\tau}{\bar\tau}\right) = -\ln\bar\tau +\Li_2\left(-\tau/\bar\tau\right)\,,
\notag\\
\varphi_3(\tau) & = -2\varphi_2(x)-H_{101}\left(-\frac{\tau}{\bar \tau}\right)=
2H_1\left(-\frac{\tau}{\bar \tau}\right)-2H_{01}\left(-\frac{\tau}{\bar \tau}\right)-H_{101}\left(-\frac{\tau}{\bar \tau}\right)\,,
\notag\\
&= 2\ln\bar\tau -2\Li_2\left(-\tau/\bar\tau\right) - 2\Big(\Li_3(\bar \tau)-\Li_3(1)\Big)+\ln\bar\tau\left(\Li_2(\bar\tau)
+\frac{\pi^2}6\right)
+\frac16 \ln^3\bar\tau
\notag\\
\varphi_4(\tau) & =-3\varphi_3(\tau)-\varphi_2(\tau) -H_{0101}\left(-\frac{\tau}{\bar \tau}\right)
\notag\\
&=
-5H_1\left(-\frac{\tau}{\bar \tau}\right) + 5H_{01}\left(-\frac{\tau}{\bar \tau}\right)+3H_{101}\left(-\frac{\tau}{\bar \tau}\right)
-H_{0101}\left(-\frac{\tau}{\bar \tau}\right),
\label{phi-tau}
\end{align}
where $H_{p\ldots q}(z)$ are harmonic polylogarithms, see Ref.~\cite{Remiddi:1999ew}.

With these expressions at hand, our result for the invariant kernel~\eqref{Hinv3} is complete. We obtain
\begin{align}\label{tau-x}
 \chi_{\rm inv}^{(3)}(\tau) &= \sum_{k=1}^{4} B^{(3+)}_k \varphi_k(\tau) - C^{(3+)}_0 + C^{(3+)}_1 \big[\ln\left(\tau/\bar\tau\right)+2\big] + \delta {\chi}^{(3+)}_{\rm inv}(\tau)\,,
\notag\\
 \chi_{\rm inv}^{\mathbb{P}(3)}(\tau) &= \sum_{k=1}^{4} B^{(3-)}_k \varphi_k(\tau) - C^{(3-)}_0 + \delta {\chi}^{(3-)}_{\rm inv}(\tau)\,,
\end{align}
where the functions $\varphi_k(\tau)$ are defined in Eq.~\eqref{phi-tau}, the coefficients $B^{(3\pm)}_k$, $C^{(3\pm)}_k$ can be found in Table~\ref{tab:resultcoeff} and
the parameters for  $\delta {\chi}^{(3\pm)}_{\rm inv}(\tau)$ \eqref{deltachi-tau} are collected in Table~\ref{tab:fitcoeff}.

\begin{figure}[t]
\begin{center}
\scalebox{0.6}{
\begin{picture}(300,270)(0,0)
\put(-5,0){\epsfxsize11cm\epsffile{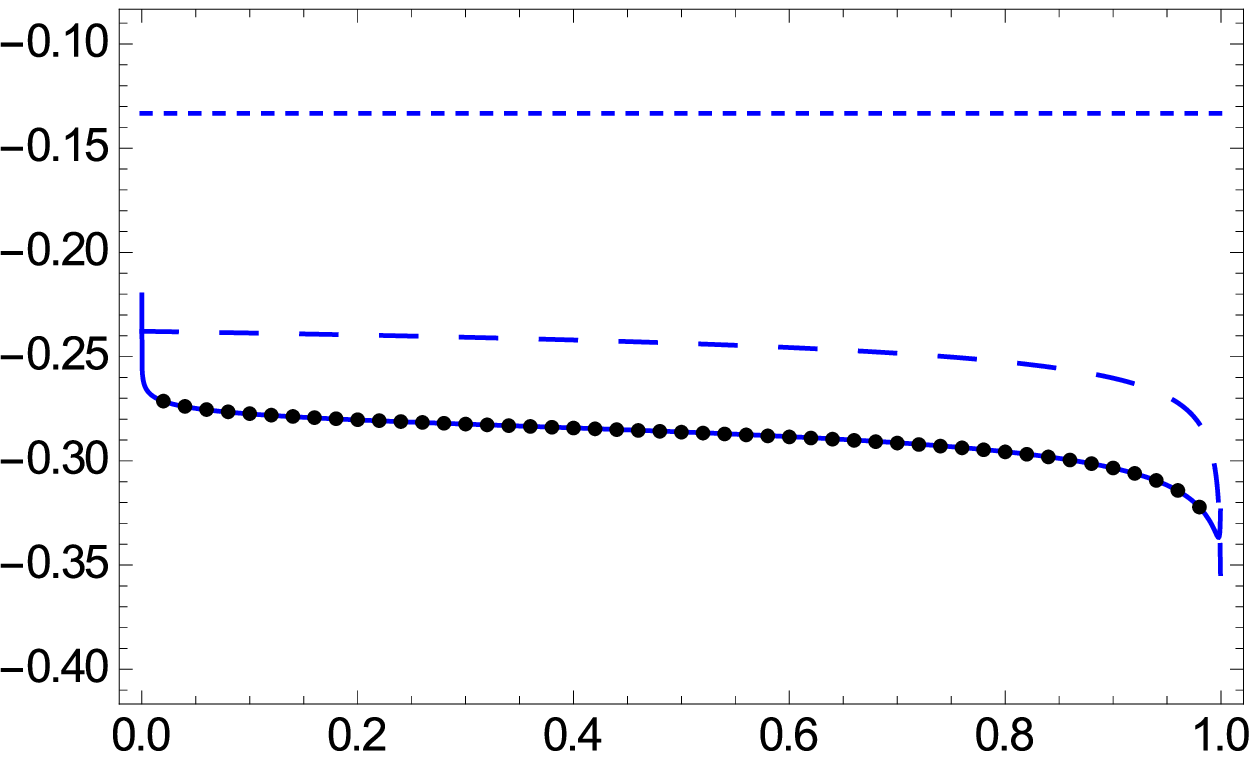}}
\put(40,30){\LARGE{$ \chi_{\rm inv}(\tau ) $}}
\put(160,0){\LARGE{$ \tau $}}
\end{picture}
}
\hspace{1cm}
\scalebox{0.6}{
\begin{picture}(300,270)(0,0)
\put(-5,0){\epsfxsize11cm\epsffile{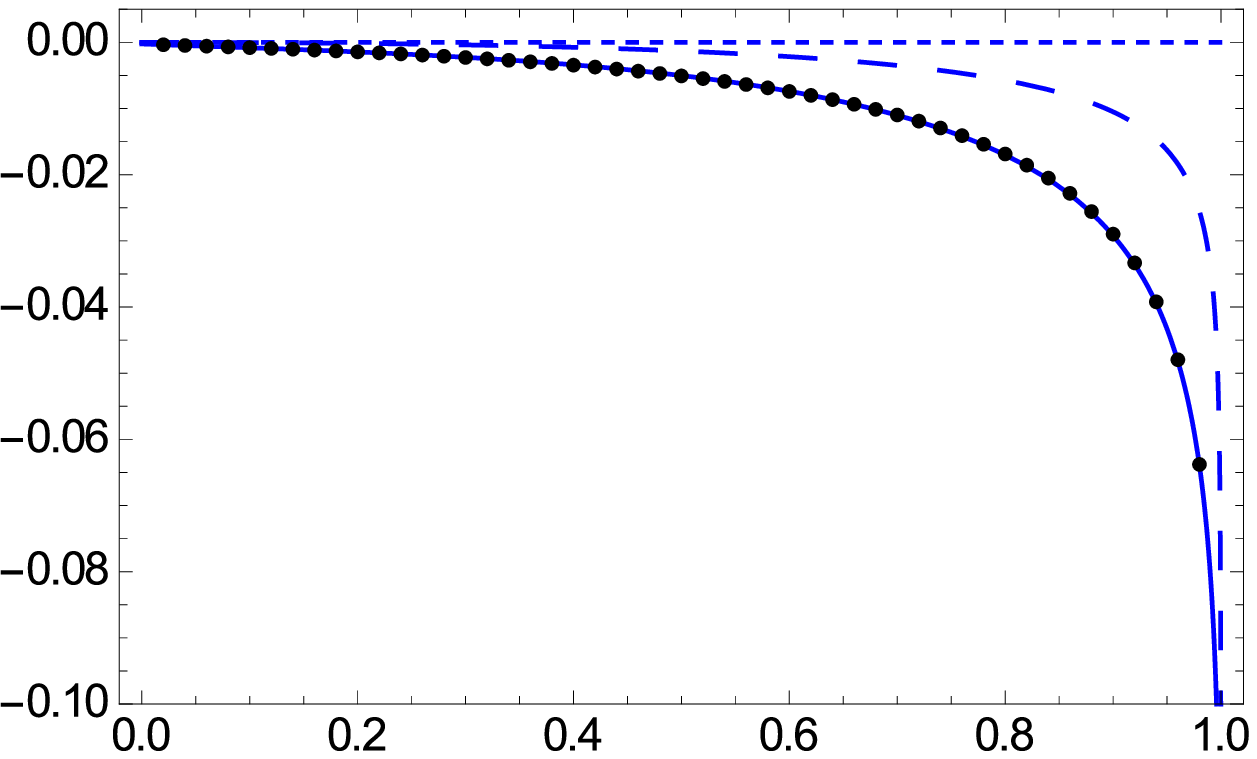}}
\put(40,30){\LARGE{$ \chi_{\rm inv}^{\mathbb P}(\tau ) $}}
\put(160,0){\LARGE{$ \tau $}}
\end{picture}
}
\end{center}
\caption{\small Invariant functions $\chi_{\rm inv}(\tau)$ (left panel) and
$\chi_{\rm inv}^{\mathbb P}(\tau )$ (right panel) for $\alpha_s/\pi=0.1$.
The LO result (short dashes) is shown together with the NLO (long dashes)
and NNLO (solid curves). The NNLO results using exact $\mathcal{O}(a^3)$ functions obtained
by the numerical integration of  Eq.~\eqref{Legendre} are shown by black dots
for comparison.
}
\label{fig:chi}
\end{figure}

For illustration, in Fig. \ref{fig:chi} we compare the full NNLO invariant functions $\chi(a) = a \chi^{(1)} + a^2  \chi^{(2)} + a^3 \chi^{(3)}$
with the NLO, $\mathcal{O}(a^2)$, and the LO, $\mathcal{O}(a)$, results for a typical value of the coupling $\alpha_s/\pi = 0.1$ and,
for definiteness, $n_f=4$. In the same plot the NNLO results using the exact three-loop functions obtained
by the numerical integration of Eq.~\eqref{Legendre} are shown by dots. One sees that the accuracy of our parametrization is
rather good. The remaining entries in the invariant kernel are, for the same values $n_f=4$ and $\alpha_s/\pi = 0.1$,
\begin{align}
  \Gamma_{\rm cusp} &= a \Gamma^{(1)} (1+8.019 a+ 80.53 a^2 +\ldots)  = a \Gamma^{(1)} (1+ 0.2005 + 0.0503 +\ldots)  \,,
\notag\\
  \chi_0 &= a \chi_0^{(1)} (1 -0.7935 a - 141.3 a^2+\ldots)   = a \chi_0^{(1)} (1 - 0.0198  - 0.0883 +\ldots) \,.
\end{align}

%

%
\section{From light-ray to local operators}\label{sec:localOPE}
%

Light-ray operators are nothing but the generating functions for the renormalized local operators
so that the mixing matrices for flavor-nonsinglet local operators can be calculated, in principle, by
evaluating the evolution kernels on the test functions of the form $f(z_1,z_2) = z_1^{m} z_2^k$, cf.~\eqref{LRO}.
The results in this form are required for several applications, e.g.
the calculation of moments of the distribution amplitudes and generalized nucleon parton distributions using
lattice QCD techniques where the precision is increasing steadily and in some
case already now requires NNLO accuracy~\cite{Braun:2015axa}.

Instead of using mixing matrices for the operators with a given number of left and right derivatives, as in Eq.~\eqref{LRO},
it proves to be more convenient to go over to the Gegenbauer polynomial basis.
To the leading-order accuracy these operators diagonalize the evolution equations.
Apart from convenience, writing the results in this form will allow us to make explicit connection to the formalism
and notations used in~\cite{Belitsky:1998gc} where the NLO expressions 
have been presented in this basis.
We will see that the local operator formalism has its own advantages, e.g., solving the conformal
constraint \eqref{HDelta} is significantly easier in this language.
Also the final step, reconstructing the invariant kernels from the eigenvalues,
can be completely avoided here, as they directly enter the anomalous dimension matrices as diagonal elements.

Our goal is to translate the evolution kernels for light-ray operators
into the anomalous dimension matrices for local operators of the form
\begin{align}
\mathcal O_{nk}
&= (\partial_{z_1}+\partial_{z_2})^k C_n^{3/2} \left( \frac{\partial_{z_1}-\partial_{z_2}}{\partial_{z_1}+\partial_{z_2}} \right) \mathcal O(z_1,z_2)\bigg| _{z_1 = z_2 = 0}\,,\qquad k\ge n\,.
\label{Onk}
\end{align}
Here $k$ is the total number of derivatives and the operator of the lowest dimension for given $n$, 
$\mathcal O_n\equiv \mathcal O_{nn}$, is a conformal operator
(lowest weight of the representation of the $SL(2)$ group). Increasing $k$ for fixed $n$ corresponds to adding total derivatives.

The operators $\mathcal O_{nk}$ mix under the evolution
\begin{align}
\left(\mu\frac{\partial}{\partial\mu} + \beta(a) \frac{\partial}{\partial a} \right) [\mathcal O_{nk}] = - \sum_{n'=0}^n \gamma_{nn'}\, [\mathcal O_{n'k}]\,.
\end{align}
The mixing matrix $\gamma_{nn'}$ is triangular and its diagonal elements are equal to the anomalous dimensions
\begin{align}
 \gamma_{nn'}  = 0\quad\text{if}\quad n'>n\,, && \gamma_{nn} = \gamma_{n}\,.
\end{align}
Since $\gamma_{nn'}$ does not depend on $k$, the second subscript $k$ is essentially redundant.
In what follows we will use a ``hat'' for the anomalous dimensions and other quantities in matrix notation
\begin{align}
  \widehat{\boldsymbol{\gamma}} \equiv \gamma_{nn'}\,.
\end{align}

The light-ray operator~\eqref{LRO} can be expanded in terms of the local operators defined in Eq.~\eqref{Onk}
\begin{align} \label{localExpand}
  \mathcal O(x; z_1,z_2 ) = \sum _{n=0}^\infty\sum_{k=n}^{\infty} \Phi_{nk}(z_1,z_2) \mathcal O_{nk}(x) \,,
\end{align}
where the coefficients $\Phi_{nk}(z_1,z_2)$ are homogeneous polynomials of two variables of degree $k$~\cite{Braun:2011dg}:
\begin{align}
\Phi_{nk}(z_1,z_2) =  \omega_{nk} (S_+^{(0)})^{k-n} z_{12}^n \,,
&&
 \omega_{nk} =  2 \frac{2 n + 3}{(k-n)!} \frac{\Gamma(n + 2) }{\Gamma(n + k + 4)}\,.
\label{Phi-nk}
\end{align}
These polynomials are mutually orthogonal and form a complete set of functions%
\footnote{Our notation in this section is adapted to facilitate the comparison with Ref.~\cite{Belitsky:1998gc} and differs from the
notation used in~\cite{Braun:2011dg}. In particular the functions $\Phi_{nk}$ defined in \eqref{Phi-nk} correspond to $\Phi_{n,k-n}$ in
~\cite{Braun:2011dg}.} 
w.r.t. the canonical $SL(2)$ scalar product (see, e.g.,~\cite{Braun:2011dg})
\begin{align} \label{ortho}
  \langle \Phi_{nk} | \Phi_{n'k'} \rangle
= \delta_{kk'}\delta_{nn'}  ||\Phi_{nk}||^2 =\delta_{kk'}\delta_{nn'} \,\omega_{nk} \rho_n^{-1} \,,
&& \rho_n = \frac12{(n+1)(n+2)!}
\end{align}
so that the local operators~\eqref{Onk} can be obtained by the projection on the corresponding ``state''
\begin{align}
\mathcal O_{nk} &=   \rho_n\omega_{nk}^{-1}  \langle \Phi_{nk} |\mathcal O(z_1,z_2) \rangle = \rho_n \langle \left( S_+\right)^{k-n}z_{12}^n |\mathcal O(z_1,z_2) \rangle\,.
\end{align}
The (canonical) conformal spin generators $S^{(0)}_\pm$ act as rising and lowering operators on this space
whereas $S^{(0}_0$ is diagonal
\begin{align}\label{SPhi}
S_0^{(0)} \Phi_{nk}(z_1,z_2) =&~ (k+2)\Phi_{nk}(z_1,z_2)\,,
\notag\\
S_+^{(0)} \Phi_{nk}(z_1,z_2) =&~ (k-n+1)(n+k+4) \Phi_{nk+1}(z_1,z_2) \,,
\notag\\
S_-^{(0)} \Phi_{nk}(z_1,z_2) =&~ - \Phi_{nk-1}(z_1,z_2) \,.
\end{align}
Thus the set of coefficient functions $\Phi_{n,k}(z_1,z_2)$
for $k=\{n\ldots\infty\} $ forms an irreducible representation of the
$SL(2)$ algebra, which is usually referred to as the conformal tower.

Let~$\mathcal A$  be a certain operator~$\mathcal A$ acting on quantum fields.
Its action can be realized by the expansion in terms of local operators with ``matrix elements'' serving as expansion coefficients
\begin{align} \label{Alocal}
  [\mathcal A, \mathcal O_{nk}] = \sum _{n'k'} A_{nn'}^{kk'}\,\mathcal O_{n'k'} \,.
\end{align}
Alternatively one can represent $\mathcal A$ by some integro-differential operator~${\bf A}$
acting on the arguments $z_1,z_2$ of the light-ray operator $\mathcal O(z_1,z_2)$ and, by means of the expansion \eqref{localExpand},
on the coefficient functions,
\begin{align}\label{Anonlocal1}
  [\mathcal A, O](z_1,z_2) = \sum_{nk} \Phi_{nk}(z_1,z_2)\sum _{n'k'} A_{nn'}^{kk'}\,\mathcal O_{n'k'}
~\equiv~
  [{\bf A} O](z_1,z_2) = \sum_{nk} [{\bf A}\Phi_{nk}](z_1,z_2) O_{nk}\,.
\end{align}
Comparing the representations in Eqs.~\eqref{Alocal} and~\eqref{Anonlocal1} we see that the action of ${\bf A}$ on the coefficient functions of local operators
is given by the transposed matrix
\begin{align}
  [{\bf A}\Phi_{nk}](z_1,z_2) = \sum_{n'k'} A_{n'n}^{k'k} \Phi_{n'k'}(z_1,z_2) \,.
\end{align}
Using the orthogonality relation \eqref{ortho} one obtains
\begin{align}
  A_{nn'}^{kk'}
= ||\Phi_{nk}||^{-2}\langle \Phi_{nk}(z_1,z_2) | [{\bf A}\Phi_{n'k'}](z_1,z_2) \rangle
~\equiv~ \langle {nk} | {\bf A}|{n'k'} \rangle\,,
\end{align}
which is the desired conversion, for a generic operator, between the light-ray and local operator representations.

The ``matrix elements'' $A_{nn'}^{kk'}$ depend in general on four indices.
However, if the operator $\bf A$ has a certain (canonical) dimension, i.e $[S^{(0)}_0, {\bf A}]= d_{\bf A} \bf A$, then its matrix
elements are nonzero only if the indices satisfy the constraint
$d_{\bf A}=k-k'$. This reduces the number of independent indices by one and allows one to write
$A_{nn'}^{kk'} \equiv A_{nn'}(k)\delta_{k, k'+d_{\bf A}}$.
If, in addition,  the operator $\mathcal A$ is invariant under translations, i.e. $[S_-, {\bf A}]=0$,
then it follows from~Eqs.~\eqref{SPhi} that the matrix elements $A_{nn'}^{kk'} \equiv A_{nn'}$ do not depend on
the upper indices at all.

Constraints on the operator mixing in the light-ray operator representation
that follow from conformal algebra take the form~\eqref{HDelta}
\begin{align}\label{HDelta1}
\big[S_+^{(0)},\mathbb{H}(a)\big] &= \big[\mathbb{H}(a),z_1+z_2\big]\left(\bar\beta(a)+ \frac12\mathbb{H}(a)\right)+\big[\mathbb{H}(a),z_{12} \Delta(a)\big].
\end{align}
To translate this equation into the local operator representation, we define the matrices
\begin{align}
    \mathbf{a}_{mn}(k)  &= \langle m,k | S_+^{(0)}| n,k-1\rangle\,,
\notag\\
     \mathbf{b}_{mn}(k)  &= \langle m,k | z_1+z_2 | n,k-1\rangle\,,
\notag\\
   \boldsymbol{\gamma}_{mn} &= \langle m,k | \mathbb{H} | n,k\rangle\,,
\notag\\
   \mathbf{w}_{mn} &= \langle m,k | z_{12}\Delta  | n,k-1\rangle\,.
\end{align}
The first two matrix elements are easily computed,
\begin{align}
   \mathbf{a}_{mn}(k)  &= -(m-k)(m+k+3)\delta_{mn} \equiv - {\bf a}(m,k)\delta_{mn}\,,
\notag\\
   \mathbf{b}_{mn}(k)  &=  2(k-n)\delta_{mn}-2(2n+3)\vartheta_{mn} \,,
\label{ab}
\end{align}
where we introduced a discrete step function
\begin{align*}
  \vartheta_{mn} =
\begin{cases}
1 \text{ if } m - n > 0 \text{ and even} \\ 0 \text { else.}
\end{cases}
\end{align*}
The remaining two are nontrivial and can be written as a perturbative expansion
\begin{align}
 \widehat{\boldsymbol{\gamma}}(a) &= a \widehat{\boldsymbol{\gamma}}^{(1)} + a^2 \widehat{\boldsymbol{\gamma}}^{(2)} + a^3 \widehat{\boldsymbol{\gamma}}^{(3)} +\ldots\,, \qquad a= \frac{\alpha_s}{4\pi}\,.
\notag\\
 \widehat{\mathbf{w}}(a) &= a \widehat{\mathbf{w}}^{(1)} + a^2 \widehat{\mathbf{w}}^{(2)} +\ldots\,.
\end{align}

Eq.~\eqref{HDelta1} becomes in matrix notation
\begin{align} \label{GammaW}
[\widehat {\bf a},\widehat{\boldsymbol{\gamma}}(a) ]
= [\widehat{\boldsymbol{\gamma}}(a),\widehat {\bf b}]\left(\bar \beta(a)+\frac12\widehat{\boldsymbol{\gamma}}(a)\right)
+ [\widehat{\boldsymbol{\gamma}},\widehat {\bf w}(a)]\,.
\end{align}
Note that the matrices
$\widehat {\bf a}(k)$ and ${\widehat {\bf b}}(k)$ \eqref{ab} depend in principle on the total number of derivatives $k$.
However, due to the fact that only diagonal elements depend on this parameter, the dependence on $k$ drops out in the commutator.
Hence we can safely omit it.

In complete analogy to the light-ray operator formulation, this equation fixes the non-diagonal
(i.e. canonically non-invariant) part of the anomalous dimension matrix.  Indeed, the commutator on the l.h.s. of Eq.~\eqref{GammaW}
takes the form
\begin{align}
     [\widehat {\bf a},\widehat{\boldsymbol{\gamma}}(a) ]_{mn} &= (- \mathbf{a}(m,k) + \mathbf{a}(n,k)) \gamma_{mn} = - \mathbf{a}(m,n) \gamma_{mn}\,,
\end{align}
so that the non-diagonal elements of the mixing
matrix are given by~\cite{Mueller:1993hg}
  \begin{align}
 \widehat{\boldsymbol{\gamma}}^{\text{ND}}(a) =
\mathcal G \left\{  [\widehat{\boldsymbol{\gamma}}(a),\widehat {\bf b}]\left(\frac12\widehat{\boldsymbol{\gamma}}(a)+\bar\beta(a) \right)
+ [\widehat{\boldsymbol{\gamma}}(a),\widehat {\bf w}(a)] \right\},
\label{mueller}
\end{align}
where
\begin{align}
\mathcal G \big\{\widehat M \big\}_{mn} = -\frac{M_{mn}}{ {\bf a }(m,n) } \,.
\end{align}
In particular to the two-loop accuracy 
  \begin{align}
 \widehat{\boldsymbol{\gamma}}^{(2),\text{ND}} =
\mathcal G \left\{  [\widehat{\boldsymbol{\gamma}}^{(1)},\widehat {\bf b}]\left(\frac12\widehat{\boldsymbol{\gamma}}^{(1)}+\beta_0 \right)
+ [\widehat{\boldsymbol{\gamma}}^{(1)},\widehat {\bf w}^{(1)}] \right\}.
\end{align}
Here $\widehat{\boldsymbol{\gamma}}^{(1)}$ is the well-known (diagonal) matrix of one-loop anomalous dimensions
\begin{align}
  {\gamma}_{mn}^{(1)} & = \gamma^{(1)}_n \delta_{mn} = 2 \delta_{mn} C_F \big(4S_1(n+1)-\frac{2}{(n+1)(n+2)}-3\big)
\end{align}
and $\widehat {\bf w}^{(1)}$ is the one-loop conformal anomaly
\begin{align}
  {\bf w}_{mn}^{(1)} &= 4 C_F (2n+3)\, \mathbf{a}(m,n) \left( \frac{A_{mn} - S_1(m+1)}{(n+1)(n+2)}
  + \frac{2A_{mn}}{\mathbf{a}(m,n)} \right) \vartheta_{mn}\,,
\end{align}
where
\begin{align}
 A_{mn} = S_1\left(\frac{m+n+2}2\right) -S_1\left(\frac{m-n-2}2\right) +2S_1(m-n-1)-S_1(m+1).
\end{align}
Collecting everything one obtains the two-loop anomalous dimension matrix:
\begin{align}
 \gamma^{(2)}_{mn} &= \delta_{mn}\gamma^{(2)}_n - \frac{\gamma^{(1)}_m-\gamma^{(1)}_n}{\mathbf{a}(m,n)}
\biggl\{
 - 2(2n+3) \left(\beta_0 + \frac12 \gamma^{(1)}_n\right)\vartheta_{mn}+  {\bf w}_{mn}^{(1)}
\biggr\}.
\end{align}
The first few elements $(0\leq n\leq7,~0\leq m\leq7)$ for $N_c=3$ are
\begin{align}
   {\gamma}_{mn}^{(2)} & =
\begin{pmatrix}
 0 & 0 & 0 & 0 & 0 & 0 & 0 & 0 \\
 0 & \frac{23488}{243} & 0 & 0 & 0 & 0 & 0 & 0 \\
 \frac{260}{9} & 0 & \frac{34450}{243} & 0 & 0 & 0 & 0 & 0 \\
 0 & \frac{8668}{243} & 0 & \frac{5241914}{30375} & 0 & 0 & 0 & 0 \\
 \frac{52}{9} & 0 & \frac{8512}{243} & 0 & \frac{662846}{3375} & 0 & 0 & 0 \\
 0 & \frac{120692}{8505} & 0 & \frac{261232}{7875} & 0 & \frac{83363254}{385875} & 0 & 0 \\
 -\frac{2054}{14175} & 0 & \frac{34243}{2025} & 0 & \frac{2208998}{70875} & 0 & \frac{718751707}{3087000} & 0 \\
 0 & \frac{226526}{35721} & 0 & \frac{982399}{55125} & 0 & \frac{7320742}{250047} & 0 & \frac{557098751203}{2250423000}
\end{pmatrix}
\notag
\\[4mm] &\hspace*{1cm}
 -n_f
\begin{pmatrix}
 0 & 0 & 0 & 0 & 0 & 0 & 0 & 0 \\
 0 & \frac{512}{81} & 0 & 0 & 0 & 0 & 0 & 0 \\
 \frac{40}{9} & 0 & \frac{830}{81} & 0 & 0 & 0 & 0 & 0 \\
 0 & \frac{88}{27} & 0 & \frac{26542}{2025} & 0 & 0 & 0 & 0 \\
 \frac{104}{45} & 0 & \frac{1064}{405} & 0 & \frac{31132}{2025} & 0 & 0 & 0 \\
 0 & \frac{1144}{567} & 0 & \frac{232}{105} & 0 & \frac{1712476}{99225} & 0 & 0 \\
 \frac{4108}{2835} & 0 & \frac{242}{135} & 0 & \frac{1804}{945} & 0 & \frac{3745727}{198450} & 0 \\
 0 & \frac{2372}{1701} & 0 & \frac{506}{315} & 0 & \frac{2860}{1701} & 0 & \frac{36241943}{1786050} \\
\end{pmatrix}.
\end{align}
Our expressions for the one-loop conformal anomaly and the two-loop anomalous dimension matrix coincide identically with the
results in~\cite{Belitsky:1998gc}.%
\footnote{The explicit relation to the notations in \cite{Belitsky:1998gc} is as follows:
$\widehat {\mathbf{a}}(k)|_{\text{\cite{Belitsky:1998gc}}} = -\frac12 \widehat {\mathbf{a}}(k)$,
$\widehat {\mathbf{b}}(k)|_{\text{\cite{Belitsky:1998gc}}} = \widehat {\mathbf{b}}(k)$,
$\widehat {\mathbf{w}}^{(1)}|_{\text{\cite{Belitsky:1998gc}}} = - \widehat {\mathbf{w}}^{(1)}$,
and  $\widehat{\gamma}^{(i)}_{\text{\cite{Belitsky:1998gc}}} = 2^i \widehat\gamma^{(i-1)}$.
A perturbative expansion in~\cite{Belitsky:1998gc} is done in powers of $\alpha_s/(2\pi)$, e.g.,
$\widehat{\boldsymbol{\gamma}}(a) = \sum_i (2a)^i\widehat{\boldsymbol{\gamma}}^{(i-1)} $.}

Expanding Eq.~\eqref{mueller} to the third order, we obtain the three-loop nondiagonal anomalous dimension matrix
in the form
\begin{align}
 \widehat{\boldsymbol{\gamma}}^{(3),\text{ND}} =
 \mathcal G \bigg\{
      [\widehat{\boldsymbol{\gamma}}^{(2)},\widehat{\bf b}](\frac12\widehat{\boldsymbol{\gamma}}^{(1)}+\beta_0)
+ [\widehat{\boldsymbol{\gamma}}^{(2)},\widehat{\bf w}^{(1)}]+
      [\widehat{\boldsymbol{\gamma}}^{(1)},\widehat{\bf  b}](\frac12\widehat{\boldsymbol{\gamma}}^{(2)}+\beta_1)
+ [\widehat{\boldsymbol{\gamma}}^{(1)},\widehat{\bf  w}^{(2)}]
 \bigg\}\,.
\end{align}
In addition to the already known quantities, this expression involves the matrix element of the two-loop conformal
anomaly~\eqref{Delta2plus},
\begin{align}
  \mathbf{w}_{mn}^{(2)}  =  \langle m,k| z_{12} \Delta^{(2)} | n,k-1 \rangle =  \frac14 [\widehat{\boldsymbol{\gamma}}^{(2)}, \widehat{\mathbf{b}}]  + \Delta {\bf w}_{mn}^{(2)}
\end{align}
where
\begin{align}
  \Delta {\bf w}_{mn}^{(2)} =  \langle m,k| z_{12} \Delta_+^{(2)} | n,k-1 \rangle \,.
\end{align}
The explicit expression for the operator $\Delta_+^{(2)}$ \eqref{Delta+twoloops} can be found in~\cite{Braun:2016qlg}.
We have not found a closed analytic expression for the matrix $\Delta {\bf w}_{mn}^{(2)}$, but
the values for given $m,n$ can be evaluated in a straightforward way.

Splitting the result into different color structures,
\begin{align*}
  \Delta \widehat{\bf w}^{(2)} =
C_F^2\Delta \widehat{\bf w}^P + \frac{C_F}{N_c} \Delta \widehat{\bf w}^{FA}
+ \beta_0 C_F \Delta \widehat{\bf w}^{bF},
\end{align*}
we get for the first few elements $(0\leq n \leq 5 ,~1  \leq m \leq 7)$
\begin{align}
\Delta \widehat{\bf  w}^{FA}=&
\begin{pmatrix}
 0 & 0 & 0 & 0 & 0 & 0\\
 -\frac{75}{4} & 0 & 0 & 0 & 0 & 0 \\
 0 & -\frac{5075}{108} & 0 & 0 & 0 & 0\\
 -\frac{679}{15} & 0 & -\frac{58723}{720} & 0 & 0 & 0\\
 0 & -\frac{7399}{90} & 0 & -\frac{724339}{6000} & 0 & 0\\
 -\frac{1070777}{16800} & 0 & -\frac{12001}{96} & 0 & -\frac{123357091}{756000}
   & 0 \\
 0 & -\frac{22974677}{211680} & 0 & -\frac{101507627}{588000} & 0 &
   -\frac{308384869}{1481760} \\
\end{pmatrix},
\notag\\[2mm]
\Delta \widehat{\bf  w}^{P}=&
\begin{pmatrix}
 0 & 0 & 0 & 0 & 0 & 0 \\
 -\frac{2965}{144} & 0 & 0 & 0 & 0 & 0 \\
 0 & -\frac{1176553}{10800} & 0 & 0 & 0 & 0 \\
 -\frac{140959}{9000} & 0 & -\frac{7387709}{36000} & 0 & 0 & 0\\
 0 & -\frac{75208391}{617400} & 0 & -\frac{2111899581}{6860000} & 0 & 0\\
 -\frac{68372343}{5488000} & 0 & -\frac{5045910661}{21168000} & 0 &
   -\frac{307457793929}{740880000} & 0\\
 0 & -\frac{99911324293}{800150400} & 0 & -\frac{808931234579}{2222640000} & 0 &
 -\frac{2942615103467}{5601052800} \\
\end{pmatrix},
\notag\\[2mm]
\Delta \widehat{\bf  w}^{bF}=&
\begin{pmatrix}
 0 & 0 & 0 & 0 & 0 & 0 \\
 35 & 0 & 0 & 0 & 0 & 0 \\
 0 & \frac{1855}{27} & 0 & 0 & 0 & 0\\
 \frac{105}{2} & 0 & \frac{2555}{24} & 0 & 0 & 0 \\
 0 & \frac{2891}{30} & 0 & \frac{146839}{1000} & 0 & 0 \\
 \frac{6459}{100} & 0 & \frac{390313}{2700} & 0 & \frac{2552407}{13500} & 0 \\
 0 & \frac{202829}{1764} & 0 & \frac{4798313}{24500} & 0 &
   \frac{14365013}{61740} \\
\end{pmatrix}.
\end{align}
Using these expressions and the diagonal matrix elements from~\cite{Moch:2004pa} we obtain the full three-loop
anomalous dimension matrix
\begin{align}
\widehat{\boldsymbol{\gamma}}^{(3)} =
{\rm diag}\{\gamma_0^{(3)}, \gamma_1^{(3)},\ldots \}
 + \widehat{\boldsymbol{\gamma}}^{(3)}_{\langle 1\rangle}+ n_f\, \widehat{\boldsymbol{\gamma}}^{(3)}_{\langle n_f\rangle}
+ n_f^2\, \widehat{\boldsymbol{\gamma}}^{(3)}_{\langle n_f^2\rangle} \,,
\end{align}
where the off-diagonal matrices for $N_c=3$ and different powers of $n_f$ in the range $0\leq n \leq 7 ,~0  \leq m \leq 7$
are given by the following expressions:
\begin{align}
\widehat{\boldsymbol{\gamma}}^{(3)}_{\langle 1\rangle} &=
\begin{pmatrix}
 0 & 0 & 0 & 0 & 0 & 0 & 0 & 0 \\
 0 & 0 & 0 & 0 & 0 & 0 & 0 & 0 \\
 \frac{49024}{81} & 0 & 0 & 0 & 0 & 0 & 0 & 0 \\
 0 & \frac{36623912}{54675} & 0 & 0 & 0 & 0 & 0 & 0 \\
 \frac{3911}{27} & 0 & \frac{23599891}{36450} & 0 & 0 & 0 & 0 & 0 \\
 0 & \frac{8049304723}{31255875} & 0 & \frac{320657981731}{520931250} & 0 & 0 & 0 & 0 \\
 \frac{281851388261}{7501410000} & 0 & \frac{208052194247}{714420000} & 0 & \frac{21898269506047}{37507050000} & 0 & 0 & 0 \\
 0 & \frac{7192640196053}{56710659600} & 0 & \frac{159898280729473}{525098700000} & 0 & \frac{220023775251709}{396974617200} & 0 & 0 \\
\end{pmatrix},
 \notag\\[5mm]
 \widehat{\boldsymbol{\gamma}}^{(3)}_{\langle n_f\rangle} &=
\begin{pmatrix}
 0 & 0 & 0 & 0 & 0 & 0 & 0 & 0 \\
 0 & 0 & 0 & 0 & 0 & 0 & 0 & 0 \\
 -\frac{28700}{243} & 0 & 0 & 0 & 0 & 0 & 0 & 0 \\
 0 & -\frac{5762188}{54675} & 0 & 0 & 0 & 0 & 0 & 0 \\
 -\frac{1279108}{30375} & 0 & -\frac{26434828}{273375} & 0 & 0 & 0 & 0 & 0 \\
 0 & -\frac{849255644}{18753525} & 0 & -\frac{516077668}{5788125} & 0 & 0 & 0 & 0 \\
 -\frac{54942827}{2500470} & 0 & -\frac{636248861}{13395375} & 0 & -\frac{77507831071}{937676250} & 0 & 0 & 0 \\
 0 & -\frac{1660976917}{67512690} & 0 & -\frac{7496172461}{156279375} & 0 & -\frac{36406093529}{472588830} & 0 & 0 \\
\end{pmatrix},
\notag\\[5mm]
 \widehat{\boldsymbol{\gamma}}^{(3)}_{\langle n_f^2\rangle} &=
\begin{pmatrix}
0 & 0 & 0 & 0 & 0 & 0 & 0 & 0 \\
 0 & 0 & 0 & 0 & 0 & 0 & 0 & 0 \\
 \frac{236}{81} & 0 & 0 & 0 & 0 & 0 & 0 & 0 \\
 0 & \frac{3172}{1215} & 0 & 0 & 0 & 0 & 0 & 0 \\
 \frac{1316}{2025} & 0 & \frac{41356}{18225} & 0 & 0 & 0 & 0 & 0 \\
 0 & \frac{187412}{178605} & 0 & \frac{22012}{11025} & 0 & 0 & 0 & 0 \\
 \frac{55169}{893025} & 0 & \frac{10793}{9450} & 0 & \frac{176539}{99225} & 0 & 0 & 0 \\
 0 & \frac{726193}{1607445} & 0 & \frac{681503}{595350} & 0 & \frac{515359}{321489} & 0 & 0 
\end{pmatrix}.
\end{align}
For completeness we also write the first few anomalous dimensions~\cite{Moch:2004pa}:
\begin{align}
\gamma^{(3)}_{0} = & 0 \,,
\notag\\
\gamma^{(3)}_1 = &\frac{2560}{81}\zeta_3+\frac{11028416}{6561}
-\left(\frac{2560}{27} \zeta_3 +\frac{334400}{2187}\right)n_f-\frac{1792 }{729}n_f^2 \,,
\notag\\
\gamma_2^{(3)} =&
\frac{2200}{81}\zeta_3+\frac{64486199}{26244}
-\left(\frac{4000}{27} \zeta _3 + \frac{967495}{4374}\right) n_f
-\frac{2569 }{729}n_f^2 \,,
\notag\\
\gamma_3^{(3)} =&\frac{11512 }{405}\zeta _3+
\frac{245787905651}{82012500}
-\left(\frac{5024 }{27}\zeta _3+\frac{726591271}{2733750}\right) n_f
-\frac{384277}{91125} n_f^2 \,,
\notag\\
\gamma_4^{(3)} =&
\frac{11312}{405} \zeta _3+\frac{559048023977}{164025000}
-\left(\frac{5824}{27} \zeta _3+\frac{90842989}{303750}\right) n_f
-\frac{431242 }{91125}n_f^2 \,,
\notag\\
\gamma_5^{(3)} =&
\frac{558896}{19845} \zeta _3+\frac{10337334685136687}{2756768175000}
-\left(\frac{45376}{189} \zeta _3+\frac{713810332943}{2187911250}\right) n_f
-\frac{160695142}{31255875} n_f^2 \,,
\notag\\
\gamma_6^{(3)} =&
\frac{185482}{6615} \zeta _3+\frac{59388575317957639}{14702763600000}
-\left(\frac{16432}{63} \zeta _3+\frac{12225186887503}{35006580000}\right) n_f
-\frac{1369936511}{250047000} n_f^2 \,,
\notag\\
\gamma_7^{(3)} =&
\frac{5020814}{178605} \zeta _3 \!+\frac{46028648192099544431}{10718314664400000}
-\!\left(\!\frac{158128}{567}\zeta _3\!+\!\frac{349136571992501}{945177660000}\!\right) n_f
\!-\frac{38920977797}{6751269000} n_f^2 \,.
\end{align}
To visualize the size of the three-loop correction 
we consider the full NNLO nondiagonal part of the anomalous dimension matrix for $n_f=4$ 
in the same range $(0\leq n\leq7,~0\leq m\leq7)$: 
\begin{align}
\widehat{\boldsymbol{\gamma}}^{\rm ND}=
a^2
\begin{pmatrix}
0 & 0 & 0 & 0 & 0 & 0 & 0 & 0 \\
0 & 0 & 0 & 0 & 0 & 0 & 0 & 0 \\
11.1 + 179 a & 0 & 0 & 0 & 0 & 0 & 0 & 0 \\
0 & 22.6 +290 a & 0 & 0 & 0 & 0 & 0 & 0 \\
-3.47-13.2 a & 0 & 24.5 +297 a & 0 & 0 & 0 & 0 & 0 \\
0 & 6.12 + 93.2 a & 0 & 24.3+291 a & 0 & 0 & 0 & 0 \\
-5.94 - 49.3 a & 0 & 9.74 +120 a & 0 & 23.5+282 a & 0 & 0 & 0 \\
0 & 0.0764+35.6 a & 0 & 11.4+131 a & 0 & 22.6+272 a & 0 & 0 
\end{pmatrix}.
\end{align}
For realistic values of the strong coupling $a = \alpha_s/(4\pi) \sim 1/40$ the three-loop contribution is on the average about 30\% of 
the two-loop result.

\section{Conclusions}\label{sec:summary}

Using the 
two-loop 
result for the conformal anomaly obtained in Ref.~\cite{Braun:2016qlg} 
we have completed here 
the calculation of the three-loop evolution kernel for the flavor-nonsinglet leading-twist operators in off-forward kinematics.
The result is presented in the form of the evolution equation for the relevant nonlocal light-ray operator.
In addition we derive the explicit expression for the three-loop anomalous dimension matrix for the local operators
of dimension $D \le 10$, i.e., containing up to seven covariant derivatives. In the latter form, our result is directly
applicable to the renormalization of meson distribution amplitudes and will be useful for lattice calculations
of their first few moments.

Practical methods for the solution of the three-loop evolution equations in 
more general GPD kinematics still have
to be developed. Our results can, in principle, be translated to the evolution equation for GPDs by a Fourier
transformation, although algebraic complexities of this transformation may prohibit obtaining
the analytic expressions. An alternative method using 
the Mellin transformation in 
the conformal spin~\cite{Mueller:2005ed,Kirch:2005tt} is very attractive and it has become the standard tool
in the NLO analysis of the DVCS~\cite{Mueller:2005nz,Kumericki:2006xx,Kumericki:2007sa} and deeply-virtual meson 
production~\cite{Mueller:2013caa,Duplancic:2016bge}.
The extension of this technique to NNLO was considered in \cite{Mueller:2005nz,Kumericki:2006xx,Kumericki:2007sa} in
a special ``conformal'' renormalization scheme. The transformation to the conventional $\overline{\text{MS}}$ scheme
can be done using the results of our paper, but it remains to be seen whether this works in practice
without major complications. The NNLO analysis of the DVCS data will, of course, require 
the extension of our results to flavor-singlet operators.

\section*{Acknowledgments}
\addcontentsline{toc}{section}{Acknowledgments}

This study was supported by Deutsche Forschungsgemeinschaft (DFG) with the grants 
$\text{MO~1801/1-2}$ and SFB/TRR 55.


\appendix

\section*{Appendices}
\addcontentsline{toc}{section}{Appendices}

\renewcommand{\theequation}{\Alph{section}.\arabic{equation}}
\renewcommand{\thesection}{{\Alph{section}}}
\renewcommand{\thetable}{\Alph{table}}
\setcounter{section}{0} \setcounter{table}{0}

\section{Two-loop invariant kernel}\label{app:H2loop}

The two-loop constant term  $\chi_0^{(2)}$ and the functions $\chi_{\rm inv}^{(2)}(\tau)$, $\chi_{\rm inv}^{\mathbb{P}(2)}(\tau)$ in the
invariant kernel \eqref{Hinvk} are given by the following expressions:
\begin{align}
\chi_0^{(2)} & = \frac13C_F\biggl\{ \beta_0 \big(37-4\pi^2\big) +
                         C_F\big (43-4\pi^2\big ) +\frac1{N_c}\big (26-8\pi^2+72\zeta_3\big ) \biggr\}\,,
\end{align}
and
\begin{align}
\chi_{\rm inv}^{(2)}(\tau)& = 4C_F\biggl\{ -\frac{11}3 \beta_0
+ C_F\left[\ln\bar\tau - \frac{20}3+\frac{2\pi^2}3\right]
-\frac2{N_c}\left(\Li_2(\tau)+\frac12\ln^2\bar\tau -\frac1\tau\ln\bar\tau-\frac{\pi^2}6+\frac53\right)
\biggr\}\,,
\notag\\
\chi_{\rm inv}^{\mathbb{P}(2)}(\tau) & = -\frac{4C_F}{N_c}\Big(\ln^2\bar\tau - 2 \tau \ln\bar\tau\Big)\,.
\label{chi-inv-2}
\end{align}
The corresponding kernels in Mellin space take a rather simple form
\begin{align}
\widetilde\chi_{\rm inv}^{(2)}(\rho) & = 4C_F\biggl\{  r_0 (2\pi i)\delta(-i\rho)
 -\frac{\pi}{\rho\sin\pi\rho}\left(C_F+\frac{2}{N_c}\frac{2\rho+1}{\rho(\rho+1)}\right)
\biggr\}\,,
\notag\\
\widetilde \chi_{\rm inv}^{\mathbb{P}(2)}(\rho) &=-\frac{8 C_F}{N_c}\frac{\pi}{\rho\sin\pi\rho} (\rho-1)S_1(\rho-1)\,,
\label{mellin-twoloop}
\end{align}
where
\begin{align}
r_0=-\frac{11}3 \beta_0
+ C_F\left[\frac{2\pi^2}3-\frac{20}3\right]
-\frac2{N_c}\left(\frac83-\frac{\pi^2}6\right)\,.
\end{align}
The following table of Mellin transforms in Tab.~\ref{Tab:Examples} is helpful to arrive at this representation:
\begin{table}[ht]
\centering
\begin{tabular}{|c||c|c|}
        \hline\hline
         & &\\[-1mm]
          $\chi(\tau)$ &        $-\dfrac1\pi\rho\sin(\pi\rho)\widetilde \chi(\rho)$  & $\gamma(N), \ \ n=N+1$ \\[3mm]
          \hline
          & & \\[-3mm]
           $ \ln\bar\tau $ &  $1$ & $ -\dfrac1{(n(n+1))^2}   $
           \\[3mm]
           $\dfrac{\bar\tau}\tau\ln\bar\tau+1$ &  $-\dfrac{\rho}{\rho+1}$&     $ 2S_3(n)-2\zeta_3+\dfrac1{n(n+1)} $
           \\[3mm]
            $\bar\tau\ln\bar\tau$ &  $\rho \, S_1(\rho)$ & $(-1)^{n}\biggl\{2S_{-3}(n)-4S_{-2,1}(n)
                          +2S_1(n)\left(2S_{-2}(n)+\dfrac{\pi^2}6\right)-\zeta_3\biggr\}$ \\[3mm]

        $\dfrac12\ln^2\bar\tau$   &  $ S_1(\rho-1) $ & $\dfrac{n^2+n+1}{n^3(n+1)^3}
                          -\dfrac{(-1)^n}{n(n+1)}\left(2S_{-2}(n)+\dfrac{\pi^2}6\right)$ \\[3mm]
          $\Li_2(\tau)$&  $-S_1(\rho)$ & $\dfrac{(-1)^n}{n(n+1)}\left(2S_{-2}(n)+\dfrac{\pi^2}6\right)$\\[3mm]
          \hline\hline
\end{tabular}
\caption{Mellin transformation~\eqref{mellin} for several typical kernels and the corresponding contributions to the anomalous dimensions.
Here $S_1(\rho)=\psi(\rho+1)-\psi(1)$. }
\label{Tab:Examples}
\end{table}

%
\section{$\mathbb{X}$ kernels}\label{app:X}
%
In this appendix we present explicit expressions for the kernels $\mathbb{X}^{(k)}$ appearing in the operator of
similarity transformation~\eqref{similarity1}.

The one-loop kernel $\mathbb{X}^{(1)}$ is defined as a solution to the differential equation \eqref{X1def}
\begin{align}
\big[S_+^{(0)}, \mathbb{X}^{(1)}\big] &= z_{12} \Delta^{(1)}\,,
\end{align}
where $\Delta^{(1)}$ is the $\mathcal{O}(a)$ conformal anomaly~\cite{Braun:2014vba}
\begin{align}
\Delta^{(1)} f(z_1,z_2)
&= -2C_F\int_0^1d\alpha\Big(\frac{\bar\alpha}\alpha+\ln\alpha\Big)\Big[f(z_{12}^\alpha,z_2)-f(z_1,z_{21}^\alpha)\Big]\,.
\end{align}
The result reads
\begin{align}
\mathbb{X}^{(1)} f(z_1,z_2) &= 2C_F\left(\int_0^1 d\alpha\frac{\ln\alpha}{\alpha}
\Big[2f(z_1,z_2)-f(z_{12}^\alpha,z_2)-f(z_1,z_{21}^\alpha)\Big] +\Delta \mathbb{X}^{(1)}_{\rm inv}\right)\,,
\end{align}
where  $\Delta \mathbb{X}^{(1)}_{inv}$ is an invariant kernel (solution of the corresponding homogeneous equation)
which has the following generic form (cf.~\eqref{Hinvk}):
\begin{align}
 \Delta \mathbb{X}^{(1)}_{\rm inv}f (z_1,z_2) &= X_0 f (z_1,z_2)+
X_1 \int_0^1d\alpha\frac{\bar\alpha}{\alpha} \Big(2f(z_1,z_2)-f(z_{12}^\alpha,z_2)-f(z_1,z_{21}^\alpha)\Big)
\notag\\
&\quad +\int_0^1d\alpha \int_0^{\bar\alpha} d\beta \Big(\Delta\chi(\tau)+\Delta\chi_P(\tau) \mathbb{P}_{12}\Big) f(z_{12}^\alpha,z_{21}^\beta)\,.
\label{defX1inv}
\end{align}
The choice of $\Delta \mathbb{X}^{(1)}_{inv}$ is a matter of convenience, e.g., it can be put to zero.

The two-loop kernel $\mathbb{X}^{(2)}$ is defined as a solution to Eq.~\eqref{X2def}
\begin{align}
\big[S_+^{(0)}, \mathbb{X}^{(2)}\big] &= z_{12} \Delta^{(2)} +\Big[\mathbb{X}^{(1)},z_1+z_2\Big]\left(\beta_0+\frac12\mathbb{H}^{(1)}\right)
+\frac12\Big[\mathbb{X}^{(1)},z_{12}\Delta^{(1)}\Big]\,,
\label{X2def1}
\end{align}
where~\cite{Braun:2016qlg}
\begin{align}
z_{12} \Delta^{(2)}= z_{12} \Delta^{(2)}_+ + \frac1{4}\big[\mathbb{H}^{(2)},z_1+z_2\big].
\label{Delta2plus}
\end{align}
The operator $\Delta^{(2)}_+$ takes the form
\begin{align}
  \Delta_+^{(2)} f(z_1,z_2) &=
\int_0^1\!d\alpha\!\int_0^{\bar\alpha} d\beta \Big[\omega(\alpha,\beta) + \omega^{\mathbb{P}}(\alpha,\beta) \mathbb{P}_{12}\Big]
\Big[f(z_{12}^{\alpha},z_{21}^\beta)-f(z_{12}^{\beta},z_{21}^\alpha)\Big]
\nonumber\\
&\quad
+ \int_0^1\!du\int_0^1\!dt \, \varkappa(t)\,\Big[f(z_{12}^{ut},z_2) - f(z_1,  z_{21}^{ut})\Big].
\label{Delta+twoloops}
\end{align}
Explicit expressions for the functions $\omega(\alpha,\beta)$, $\omega^{\mathbb{P}}(\alpha,\beta)$  and $\varkappa(\tau)$
can be found in~\cite{Braun:2016qlg} (see App.~C2).

The solution can be written as a sum of three terms corresponding to the three contributions on the r.h.s. of Eq.~\eqref{X2def1},
\begin{align}
     \mathbb{X}^{(2)} &= \mathbb{X}_{\rm I}^{(2)} + \mathbb{X}_{\rm II}^{(2)} + \mathbb{X}_{\rm III}^{(2)},
\end{align}
where
\begin{align}
   \mathbb{X}_{\rm I}^{(2)} &=
\frac14 \left(\mathbb{T}^{(2)}+\beta_0\mathbb{T}^{(2,1)}\right) + C_F\left(1-\frac{\pi^2}6\right)\mathbb{T}^{(1)}
+  \mathbb{X}_{\rm IA}^{(2)} +  \mathbb{X}_{\rm IB}^{(2)}\,,
\notag\\
   \mathbb{X}_{\rm II}^{(2)} &= \mathbb{X}^{(2,1)}\left(\beta_0+\frac12\mathbb{H}^{(1)}\right)\,,
\notag\\
   \mathbb{X}_{\rm III}^{(2)}&= \frac12\Big[\Delta\mathbb{X}_{inv}^{(1)},\mathbb{X}^{(1)}\Big] -\frac12 \mathbb{X}^{(2,2)}.
\end{align}
The operators $\mathbb{T}^{(2)}$, $\mathbb{T}^{(2,1)}$ are given in App.~\ref{app:T} and $\mathbb{X}^{(2,1)}$, $\mathbb{X}^{(2,2)}$
are defined as solutions to the following equations:
\begin{align}
&[S_+^{(0)},\mathbb{X}^{(2,1)}]  = [\mathbb{X}^{(1)}, z_1+z_2]\,, &&
 [S_+^{(0)},\mathbb{X}^{(2,2)}]  = [z_{12} \Delta^{(1)},\mathbb{X}^{(1)}]\,.
\end{align}
One obtains
\begin{align}
\mathbb{X}^{(2,1)} f(z_1,z_2) &= 2C_F\biggl\{- \int_0^1 d\alpha\left(\frac{\bar\alpha}{\alpha}(1-X_1)\ln\bar\alpha +\ln\alpha\right)
\Big[2 f(z_1,z_2)-f(z_{12}^\alpha, z_2)-f(z_1,z_{21}^\alpha)\Big]
\notag\\
&\quad +\int_0^1d\alpha \int_0^{\bar\alpha} d\beta\, \ln(1\!-\!\alpha\!-\!\beta)\Big(\Delta\chi(\tau)
+\Delta\chi_P(\tau) \mathbb{P}_{12}\Big) f(z_{12}^\alpha,z_{21}^\beta)\biggr\}\,,
\end{align}
where $X_1$, $\Delta\chi(\tau)$, $\chi_P(\tau)$ are the entries in \eqref{defX1inv}, and
\begin{eqnarray}
\lefteqn{\mathbb{X}^{(2,2)} f(z_1,z_2) =}
\nonumber\\&=&
 4C_F^2\biggl\{
\int_0^1d\alpha \int_0^1 du \left[\frac{\ln\bar\alpha}\alpha \left(\frac12\ln\bar\alpha+2\right)
+ \frac{\bar u}{u}\frac{\vartheta(\alpha)}{\bar\alpha}\right]
\Big[2 f(z_1,z_2)-f(z_{12}^{\alpha u}, z_2)-f(z_1,z_{21}^{\alpha u})\Big]
\nonumber\\&&
{} + \int_0^1 d\alpha\int_0^{\bar\alpha} d\beta \biggl[\frac1\tau \Big( \vartheta_+(\alpha)+\vartheta_+(\beta)\Big)
\Big[f(z_{12}^\alpha,z_{21}^\beta) - f(z_{1},z_{21}^\beta) - f(z_{12}^\alpha,z_{2}) + f(z_{1},z_{2})\Big]
\nonumber\\&&{}\hspace*{1cm}
+
\Big( \vartheta_0(\alpha) + \vartheta_0(\beta) \Big) f(z_{12}^\alpha,z_{21}^\beta)\biggr]\biggr\}\,,
\end{eqnarray}
where
\begin{align}
\vartheta_+(\alpha) & = - \frac1{\bar\alpha}\Big(\ln\alpha\ln\bar\alpha+2\alpha\ln\alpha+2\bar\alpha\ln\bar\alpha\Big)\,,
\notag\\[2mm]
\vartheta_0(\alpha) & =2\Big( \Li_3(\bar\alpha)-\Li_3(\alpha)-\ln\bar\alpha\, \Li_2(\bar\alpha)+\ln\alpha\, \Li_2(\alpha)\Big)
                                 +\frac1\alpha\ln\alpha\ln\bar\alpha
                                    +\frac{2}\alpha\ln\bar\alpha\,,
\\[2mm]
\vartheta(\alpha) &=\frac{\alpha}{\bar\alpha}\Big(\Li_2(\bar\alpha)-\ln^2\alpha\Big)-\frac12\frac{\bar\alpha}\alpha \ln^2\bar\alpha
                                 +\left(\alpha-\frac2\alpha\right)\ln\alpha\ln\bar\alpha-\left(3+\frac1{\bar\alpha}\right)\ln\alpha
                            -(\alpha-\bar\alpha)\frac{\bar\alpha}\alpha\ln\bar\alpha-2\,.
                            \notag
\end{align}
The operators $\mathbb{X}^{(2)}_{\rm IA}$, $\mathbb{X}^{(2)}_{\rm IB}$ which originate from the two-loop anomaly $\Delta^{(2)}_+$ are
rather involved. We obtain
\begin{align}
\mathbb{X}^{(2)}_{\rm IA} f (z_1,z_2)&= \int_0^1\! du\,\frac{\bar u}{u} \int_0^1\frac{d\alpha}{\bar\alpha}\,
\big[\varkappa(\alpha)-\varkappa(1)\big]\Big[2 f(z_1,z_2)-f(z_{12}^{\alpha u}, z_2)-f(z_1,z_{21}^{\alpha u})\Big]
\notag\\&\quad +
\int_0^1d\alpha\, \xi_{\rm IA}(\alpha)\Big[2 f(z_1,z_2)-f(z_{12}^{\alpha}, z_2)-f(z_1,z_{21}^{\alpha})\Big],
\end{align}
where $\varkappa(\alpha)$ is one of the functions entering the two-loop conformal anomaly~\eqref{Delta+twoloops} and
\begin{align}
\xi_{\rm IA}(\alpha)&=
2C_F^2\frac{\bar\alpha}\alpha\biggl[-\Li_3(\bar\alpha)+\ln\bar\alpha \,\Li_2(\bar\alpha)+\frac13\ln^3\bar\alpha+\Li_2(\alpha)+\frac1{\bar\alpha}\ln\alpha\ln\bar\alpha
-\frac14\ln^2\bar\alpha
\notag\\&\quad -\frac{3\alpha}{\bar\alpha}\ln\alpha-3\ln\bar\alpha\biggr]
+ \frac{C_F}{N_c} \left(\ln\alpha+\frac{\bar\alpha}{\alpha}\ln\bar\alpha\right).
\end{align}
Finally
\begin{align}
\mathbb{X}^{(2)}_{\rm IB} f (z_1,z_2)&= C_F \int_0^1d\alpha\int_0^{\bar\alpha} d\beta \biggl[C_F\, \xi_{\rm IB}^F(\alpha,\beta) +
\frac1{N_c} \Big(\xi_{\rm IB}^A \,(\alpha,\beta)
+ \xi_{\rm IB}^{A\mathbb{P}}(\alpha,\beta)\mathbb{P}_{12}\Big)\biggr] f(z_{12}^\alpha,z_{21}^\beta)\,,
\end{align}
where
\begin{align}
%
\xi_{\rm IB}^{A\mathbb{P}}(\alpha,\beta) &= -2\biggl[ \Li_3(\bar\alpha) -\Li_3\left(1-\frac\alpha{\bar\beta}\right) +
\frac1{\bar\alpha}\left[\Li_2(\alpha) -\Li_2\left(\frac\alpha{\bar\beta}\right)\right] +\ln\bar\tau
\,\Li_2\left(1-\frac\alpha{\bar\beta}\right)
\notag\\
&\quad +\frac{1+\alpha}{2\bar\alpha}\ln\bar\alpha\ln\bar\beta
+ (\alpha\leftrightarrow \beta) \biggr]\,,
\notag\\[1mm]
%
\xi_{\rm IB}^{A}(\alpha,\beta)&=2\biggl[\Li_3(\bar\beta) -2\Li_3(\beta)-\Li_3\left(1-\frac{\beta}{\bar\alpha}\right)-\Li_3\left(\frac{\beta}{\bar\alpha}\right)
 +\ln\bar \tau\,\Li_2\left(1-\frac{\beta}{\bar\alpha}\right)
+\ln \left(\frac\beta{\bar\alpha}\right) \Li_2(\beta)
\notag\\
&
\quad
+\frac1\alpha\left( \Li_2(\beta)-\Li_2\left(\frac{\beta}{\bar\alpha}\right)\right)
+2\frac{\bar\beta}\beta \Li_2(\beta)+\frac12 \ln\bar\alpha\ln\bar\beta +\frac{\bar\beta}\beta \ln\beta\ln\bar\beta
+(\alpha\leftrightarrow \beta) \biggr]
\end{align}
and
\begin{align}
\xi_{\rm IB}^{F}(\alpha,\beta) & = \ln(1-\alpha-\beta)\ln (\tau\,\bar\tau)
- \frac13 \ln^3(1-\alpha-\beta)
+ 3\ln\bar\alpha\ln\bar\beta -\ln\alpha\ln\beta
\notag\\
&\quad
+\biggl[-6\Li_3(\bar\alpha)-10\Li_3(\alpha) + 2\ln\bar\alpha\Li_2(\bar\alpha)
                 +6\ln\alpha\Li_2(\alpha) + \ln\alpha\ln\bar\alpha\big(\ln\alpha+\ln\bar\alpha-2\big)
\notag\\
&\quad
-4 \frac{1+\alpha}\alpha\Big(\Li_2(\bar\alpha)-\Li_2(1)\Big) - \frac13 \ln^3\bar\alpha - \frac{\bar\alpha}\alpha \ln^2\bar\alpha + \frac12\ln^2\alpha
-\frac2{\bar\alpha}\ln\alpha + \frac4{\alpha}\ln\bar\alpha +15\ln\bar\alpha
\notag\\
&\quad
+(\alpha\leftrightarrow\beta)\biggr].
\end{align}
In all expressions $\tau = \frac{\alpha\beta}{\bar\alpha\bar\beta}$.

%
\section{$\mathbb{T}$ kernels}\label{app:T}
%

The $\mathbb{T}^{(k)}$ operators, $k=1,2,\ldots$ are defined as solutions to the differential equation~\eqref{T-eq}
\begin{align}
 [S_+^{(0)},\mathbb{T}^{(k)}]   &= [\mathbf{H}_{\rm inv}^{(k)},z_1+z_2],
\end{align}
where $\mathbf{H}_{\rm inv}^{(k)}$ are the $SL(2)$ invariant parts of the evolution kernel which have the
general decomposition  as shown in Eq.~\eqref{Hinvk}. This equation can easily be solved:
\begin{align}\label{Tk}
\mathbb{T}^{(k)} f (z_1,z_2) &=-
\Gamma_{\rm cusp}^{(k)} \int_0^1d\alpha\frac{\bar\alpha \ln\bar\alpha}{\alpha} \Big(f(z_{12}^\alpha,z_2)+f(z_1,z_{21}^\alpha)\Big)
\notag\\
&\quad +\int_0^1d\alpha \int_0^{\bar\alpha} d\beta \ln(1-\alpha-\beta) \Big(\chi_{\rm inv}^{(k)}(\tau)+\chi_{\rm inv}^{\mathbb{P}(k)}(\tau)\mathbb{P}_{12}\Big) f(z_{12}^\alpha,z_{21}^\beta)\,,
\end{align}
where the functions $\chi_{\rm inv}^{(k)}$, $\chi_{\rm inv}^{\mathbb{P}(k)}$ for one loop, $k=1$, and two loops, $k=2$,
are given in Eqs.~\eqref{chi-inv-1} and \eqref{chi-inv-2}, respectively.

The $\mathbb{T}^{(2)}_1$ kernel is defined as the solution to
\begin{align}
[S_+^{(0)},\mathbb{T}^{(2)}_1] =[\mathbb{T}^{(1)},z_1+z_2]\,.
\end{align}
Using the explicit expression for $\mathbb{T}^{(1)} = \mathbf{T}_{\rm inv}^{(1)}$~\eqref{Honeloop} one obtains
after a short calculation
\begin{align}\label{T21}
\mathbb{T}^{(2)}_1 f (z_1,z_2) &=-\frac12
 \int_0^1\!d\alpha\frac{\bar\alpha \ln^2\bar\alpha}{\alpha} \Big(f(z_{12}^\alpha,z_2)+f(z_1,z_{21}^\alpha)\Big)
+\frac12\int_0^1\!d\alpha \int_0^{\bar\alpha}\! d\beta \ln^2(1\!-\!\alpha\!-\!\beta) f(z_{12}^\alpha,z_{21}^\beta)\,.
\end{align}

We stress again that Eqs.~\eqref{T-eq} determine the $\mathbb{T}$-kernels up to the
$SL(2)$ (canonically) invariant parts. Our choice in \eqref{Tk} and \eqref{T21}
corresponds to the following condition on the eigenvalues of these kernels:  Let $\gamma_{\rm inv}^{(k)}(N)$ be the
eigenvalues of the invariant kernels $\mathbf{H}_{\rm inv}^{(k)}$
(corresponding contributions to the anomalous dimensions),
\begin{align}
 \mathbf{H}_{\rm inv}^{(k)} z_{12}^N &= \gamma_{\rm inv}^{(k)}(N) z_{12}^N\,.
\end{align}
It is easy to check that eigenvalues of the kernels defined in \eqref{Tk} and \eqref{T21}
are given by the following expressions:
\begin{align}
 \mathbb{T}^{(k)} z_{12}^N & = {T}^{(k)}(N) z_{12}^N\,, \qquad  {T}^{(k)}(N) = \frac{d}{dN}\gamma_{\rm inv}^{(k)}(N)\,,
\notag\\
 \mathbb{T}_1^{(2)} z_{12}^N & = {T}_1^{(2)}(N) z_{12}^N\,, \qquad  {T}_1^{(2)}(N) = \frac12\frac{d^2}{dN^2}\gamma_{\rm inv}^{(1)}(N)\,.
\label{adim-T}
\end{align}
This choice is convenient for our present purposes as it leads to a certain symmetry of the three-loop
invariant kernel $\mathbf{H}_{\rm inv}^{(3)}$ that allows one to obtain somewhat simpler expressions, see Sec.~\ref{sec:H3inv}.


\end{document}